\documentclass[prd,aps,eqsecnum,floatfix,nofootinbib,preprint,tightenlines]{revtex4}

\usepackage{latexsym}
\usepackage{graphicx}
\usepackage{psfrag}
\usepackage{amsmath,amssymb,dsfont}
\usepackage[dvipsnames]{xcolor}
\usepackage{float}
\def\bibi{\bibitem}

\let\inodot=\i

\def\a{\alpha}
\def\b{\beta}

\def\d{\delta}
\def\e{\epsilon}                % Also, \varepsilon
\def\g{\gamma}

\def\i{\iota}

\def\l{\lambda}
\def\m{\mu}
\def\n{\nu}
\def\o{\omega}
\def\p{\pi}                     % Also, \varpi
\def\r{\rho}                    %       \varrho
\def\s{\sigma}                  %       \varsigma
\def\t{\tau}

\def\D{\Delta}

\def\G{\Gamma}

\def\P{\Pi}

\def\co{{\cal O}}

\def\cbo{{\,\raise-.15ex\Sc [\,}}                       % curly "
\def\ddt#1{{\buildrel {\hbox{\LARGE .\kern-2pt.}} \over {#1}}}% double dot-over
\def\ie{\mbox{\it i.e.}}

\def\half{{1\over 2}}
\def\Re{{\rm Re\,}}

\long \def \blockcomment #1\endcomment{}
\def\ttl#1{{\it #1}}
\def\floatcaption#1#2{ \caption{ #2 \ [#1] \label{#1}} }
\def\floatcaption#1#2{ \caption{#2 \label{#1}} }

\def\bP{{\overline{\P}}}

\begin{document}

\vspace{-2cm}
\begin{center}
\begin{boldmath}
{\Large \bf Comparison of the hadronic vacuum polarization between 
hadronic $\t$-decay data and lattice QCD}\\[1cm]
\end{boldmath}

Noah Allen,$^a$ Diogo Boito,$^b$ Maarten Golterman,$^{a,c}$
Kim Maltman,$^{d,e}$\\
Lucas M. Mansur,$^f$ and Santiago Peris$^c$\\[1cm]

$^a$Department of Physics and Astronomy, San Francisco State University\\
San Francisco, CA 94132, USA\\
[5mm]
$^b$Instituto de F{\'\inodot}sica de S\~ao Carlos, Universidade de S\~ao Paulo,
13566-590\\ S\~ao Carlos, SP, Brazil\\
[5mm]
$^c$Department of Physics and IFAE-BIST, Universitat Aut\`onoma de Barcelona\\
E-08193 Bellaterra, Barcelona, Spain\\
[5mm]
$^d$Department of Mathematics and Statistics,
York University\\  Toronto, ON Canada M3J~1P3
\\[5mm]
$^e$CSSM, University of Adelaide, Adelaide, SA~5005 Australia
\\[5mm]
$^f$Institut f\"ur Kernphysik, Johannes Gutenberg-Universit\"at Mainz\\ D-55128 Mainz, 
Germany
\\[10mm]

\end{center}

\begin{quotation}
We compare the isospin-one, vector-current hadronic vacuum polarization (HVP) 
obtained from isospin-symmetric lattice QCD with that obtained from a 
dispersive representation employing inclusive hadronic $\tau$ decay 
data corrected for isospin breaking. We consider the subtracted 
HVP evaluated at squared Euclidean momenta ranging from 
$0.5$~GeV$^2$~to $12$~GeV$^2$, together with the light-quark-connected HVP contribution
to the muon anomalous magnetic moment and the short-, intermediate- and 
long-distance RBC/UKQCD window components thereof. Dispersive 
contributions from the region of hadronic invariant masses above the $\tau$~mass are 
evaluated using perturbative QCD. We also consider dispersive 
determinations using $\tau$ data only for contributions from two-pion, 
or two-pion and four-pion, modes, and evaluating the remaining contributions 
using exclusive-mode $e^+e^-\to\mbox{hadrons}$ cross sections up 
to about~2~GeV, lessening the dependence on perturbation theory. We find 
generally good agreement between lattice and $\tau$-based results.  However, a 
comparison of $\tau$-based window-quantity contributions for 
the two four-pion modes to expectations for those contributions based on the
Pais relations and $e^+e^-$ four-pion cross sections, reveals 
significant differences for the $2\pi^-\pi^+\pi^0$ mode.

\end{quotation}

\section{\label{intro} Introduction}
Because of the recent high-precision measurement of the anomalous magnetic 
moment of the muon $a_\m=(g-2)/2$ \cite{Muong-2:2006rrc,Muong-2:2025xyk}, 
there has been much interest in the hadronic
vacuum polarization (HVP) contribution $a_\m^{\rm HVP}$, which plays a crucial 
role in the
determination of the Standard Model (SM) expectation for $a_\m$.  The 
reason is that, at present, the error on $a_\m^{\rm HVP}$ strongly
dominates the total error on the SM expectation \cite{WP25}.   Moreover, there are 
large discrepancies between the
estimate for $a_\m^{\rm HVP}$ from lattice QCD and dispersive estimates obtained 
from the 
$R$-ratio measured in $e^+e^-\to\mbox{hadrons}(\g)$ \cite{WP25}, with the 
exception of those using CMD-3 results for the two-pion exclusive-mode 
contribution~\cite{CMD3}.

These discrepancies can be traced to the exclusive electroproduction mode
$e^+e^-\to\p^+\p^-$, for which there are significant differences in the cross
section in the region around the $\r$ peak.   Because this exclusive mode has 
isospin 1, 
data for the dispersive analysis can also be obtained from the decay 
$\t^-\to\p^-\p^0\n_\t$, if one can 
reliably correct for the differences in isospin breaking (IB) between the two 
processes.   Much effort has gone into this approach; for
a partial list of references, see 
Refs.~\cite{Davier:2023fpl,Davier:2010fmf,Miranda:2020wdg,Masjuan:2023qsp,
Castro:2024prg,Colangelo:2022prz,Hoferichter:2023sli,Davier:2025jiq,CCHH} and 
other references therein.

In this paper, while also considering
the question of to what extent information from hadronic $\t$ decays can 
help provide additional insight, we take a slightly different approach, 
focusing on the facts that (i)  the $I=1$ contribution
to $a_\m^{\rm HVP}$ can be obtained from the inclusive vector isovector spectral 
function
obtained from non-strange vector hadronic $\t$ decays, and (ii) in
the isospin limit, the $I=1$ contribution to $a_\m^{\rm HVP}$ is directly related to the
light-quark connected (lqc) contribution computed in isospin-symmetric lattice QCD:
\begin{equation}
\label{I1lqc}
a_\m^{\rm HVP,lqc}=\frac{10}{9}\,a_\m^{{\rm HVP},I=1}\ .
\end{equation}
Our primary goal is to use 
hadronic $\t$-decay data in the vector isovector channel, and 
to compare results based on these data with results obtained from lattice QCD.
The quantities we are interested in are  $a_\m^{\rm HVP,lqc}$, the 
lqc parts of the 
window quantities defined in Ref.~\cite{RBC18}, and
values for the $I=1$ component of the subtracted vacuum polarization $\bP(Q^2)$ 
as a function of
Euclidean momentum $Q$ which contributes to the running of the 
electromagnetic (EM) coupling
$\a$, and for which values were obtained in lattice QCD \cite{Mainz25,BMW20}. 
Results for the lqc
intermediate window are of particular interest given that the largest of 
the discrepancies between pre-CMD-3 dispersive and lattice results
was observed for this quantity~\cite{Benton:2023dci}.

Most lattice computations of $a_\m^{\rm HVP,lqc}$ are carried out in the 
``time-momentum''
representation \cite{BM}, in which $a_\m^{\rm HVP,lqc}$ is expressed as a weighted
integral over the zero-spatial-momentum current-current 
correlator $C^{I=1}(t)$, where $t$ is the Euclidean time.   This correlator can be obtained from a Laplace transform 
of the inclusive vector isovector
spectral function, 
$\rho_V^{I=1}(s)$, where $s$ is the hadronic invariant mass-squared.   For 
squared 
invariant masses $s\le m_\tau^2$, $\rho_V^{I=1}(s)$ can be obtained from the 
differential distributions measured in non-strange hadronic $\tau$ decays. For 
$s>m_\tau^2$, where no such experimental data are available, we determine 
$\rho_V^{I=1}(s)$ using perturbation theory (PT) to fifth order in the strong 
coupling $\a_s(m_\t^2)$ \cite{PT},\footnote{Only an estimate of the 
fifth-order term is known, see Eq.~(\ref{cs}).} employing a model to estimate 
uncertainties due to residual quark-hadron duality violations in 
that region.

For the $\tau$-based input for $\rho_V^{I=1}(s)$ in the region 
$s\le m_\tau^2$, we use the results of Ref.~\cite{us25}. These were obtained by
combining $\t$-decay data from ALEPH \cite{ALEPH,ALEPH2,ALEPH13}, 
OPAL \cite{OPAL}
and Belle \cite{Belle} for the two-pion and from ALEPH and OPAL for the four-pion
contributions, with the small 
contributions from other (``residual'') modes obtained using exclusive-mode
electroproduction data (for a detailed explanation of how this was done, see
Ref.~\cite{alphas20}), with the exception of the $K^-K^0$ mode, for which $\t$-decay
data from Ref.~\cite{babarkkbartau18} was used.\footnote{A complete list of 
references used in estimating residual-mode contributions, can be found 
in Ref.~\cite{alphas20}.}   
The resulting $\rho_V^{I=1}(s)$ is the most precise vector isovector
inclusive spectral function available at present that does not involve data from
$e^+e^-$ to two or four pions.   As we will see in this paper, in addition to the 
well-known discrepancies between $\tau\to\mbox{hadrons}$-based and 
$e^+e^-\to\mbox{hadrons}$-based two-pion distributions, there 
appear to be further discrepancies between $\tau$ and electroproduction 
results for the four-pion modes.
In making comparisons between $\tau$- and electroproduction-based 
results, we will use the 2019 (``KNT19'') versions of the exclusive-mode 
electroproduction $R(s)$ contributions produced and made available to us
by the authors of Ref.~\cite{KNT19}.

The lattice results we will compare with have been computed in ``isoQCD,'' which 
is QCD without EM corrections, and in the limit $m_u=m_d$,
defined such that the pion mass is equal to the neutral pion mass.   We will 
use the definition of isoQCD specified in Ref.~\cite{FLAG}, which is sufficiently close to the
scheme used in Ref.~\cite{WP25} for our purposes.   
The inclusive $\t$-decay data we employ have already been corrected by the 
short-distance electroweak factor $S_{\rm EW}$ \cite{erlersew}, but this is not the
only source of IB in the $\t$ data, and we will need to correct the 
$\tau$-based results
for these additional IB contributions before comparing with isoQCD lattice 
results.  We note that, while EM and strong IB corrections have been computed or 
estimated by various lattice-QCD collaborations, these corrections have only 
been considered
for the EM current correlator, and not for the charged $I=1$, vector current 
case relevant for $\t$
decays.

This paper is organized as follows.   In Sec.~\ref{theory} we provide definitions and
collect the theory results we need for our analysis.   In particular, in Sec.~\ref{isospin}
we give a detailed description of our treatment of IB in the $\t$ data.   In Sec.~\ref{data}
we describe in more detail all data we employ, with emphasis on the data for 
$\t\to 2\p^-\p^+\p^0\n_\t$ and $\t\to\p^-3\p^0\n_\t$, which we compare
to Pais-relation \cite{Pais} expectations obtained 
using $e^+e^-\to 2\p^+2\p^-$ and $e^+e^-\to\p^+\p^-2\p^0$ cross-section 
input from Ref.~\cite{KNT19}. In Sec.~\ref{comp} we present our results for the
comparison between $\t$-based and lattice results, both for $\bP(Q^2)$ at a range
of values between $Q^2=0.5$ and $12$~GeV$^2$ (Sec.~\ref{vacpol}), and for 
$a_\m^{\rm HVP,lqc}$ and
the short-distance, intermediate-distance and long-distance window quantities 
(Sec.~\ref{windows}).   In Sec.~\ref{exclusive} we consider what is changed if only the
two-pion data are taken from $\t$ decays, with all other contributions taken from the
KNT19 compilation \cite{KNT19} up to the KNT19 exclusive-mode-region endpoint, 
$s=s_{\rm KNT}\equiv(1.9375\ {\rm GeV})^2$, including those of the two-pion mode
between the $\t$ mass squared and the KNT19 endpoint.   We also consider the case 
that in this approach
the four-pion data are taken from $\t$ decays instead of from the KNT19 compilation.
Section~\ref{conclusion} contains our conclusions.   Two appendices collect some 
technical details.

\section{\label{theory} Theory ingredients}
\subsection{\label{defs} Definitions}

The $I=1$ subtracted vacuum polarization $\bP^{I=1}(Q^2)$ is dispersively 
related to the 
corresponding isovector, vector spectral function $\r_V^{I=1}(s)$ as
\begin{equation}
\label{vacpolQ2}
\bP^{I=1}(Q^2)=Q^2\int_{s_{\rm th}}^\infty ds\, \frac{\r_V^{I=1}(s)}{s(s+Q^2)}\ ,
\end{equation}
where $Q$ is the Euclidean momentum, and $s$ is the invariant mass squared,
with $s_{\rm th}$ the threshold.   We decompose $\bP^{I=1}(Q^2)$ as
\begin{equation}
\label{vacpolQ2decomp}
\bP^{I=1}(Q^2)=\bP^{I=1}_{\rm data}(Q^2)+\bP^{I=1}_{\rm PT}(Q^2)+
\bP^{I=1}_{\rm DV}(Q^2)\ .
\end{equation}
This breakdown is needed because data is only available up to $s=m_\t^2$,
where $m_\t$ is the $\t$ mass, and
we thus have to rely on a theory representation for larger $s$, composed of 
a perturbation theory
(PT) part and a duality-violating (DV) part, as will be discussed below.
Likewise, we define the Euclidean time
correlation function as
\begin{equation}
\label{Ct}
C^{I=1}(t)=\half\int_{s_{\rm th}}^\infty ds\,\sqrt{s}\,e^{-\sqrt{s}t}\,\r_V^{I=1}(s)\ ,
\qquad t>0\ ,
\end{equation}
with corresponding decomposition
\begin{equation}
\label{Ctdecomp}
C^{I=1}(t)=C^{I=1}_{\rm data}(t)+C^{I=1}_{\rm PT}(t)+C^{I=1}_{\rm DV}(t)\ .
\end{equation}
This is equal to exactly $9/10$ times the light-quark-connected (lqc)
contribution to the correlation function of the electromagnetic
current, computed in Lattice QCD in the isospin limit, \ie,
\begin{equation}
\label{Ctlqc}
C^{\rm lqc}(t)=\frac{10}{9}\,C^{I=1}(t)\ .
\end{equation}
Since we will only be concerned with vector $I=1$ 
correlation functions, we will typically drop the superscript $I=1$ and 
understand the quantities of interest
to be vector, isovector channel ones in the rest of this paper.
We will, however, keep the superscript ``lqc'' where needed.

The main task of this paper is the construction of $\bP(Q^2)$ and $C(t)$ 
from the vector non-strange hadronic $\t$-decay data.   Given data
$\r(s_i)$ for the spectral function at values $s_i$ from $s_1=s_{\rm th}$ to a 
maximum value $s_N$ with $N$ the number of data points, the data-based
parts of $\bP(Q^2)$ and $C(t)$ are
\begin{eqnarray}
\label{datapart}
\bP_{\rm data}(Q^2)&=&Q^2\sum_{i=2}^N (s_i-s_{i-1})\,
\half\left(\frac{\r(s_i)}{s_i(s_i+Q^2)}+
\frac{\r(s_{i-1})}{s_{i-1}(s_{i-1}+Q^2)}\right)\ ,\\
C_{\rm data}(t)&=&\sum_{i=2}^N (s_i-s_{i-1})\,\half\left(\sqrt{s_i}\,e^{-\sqrt{s_i}t}\,
\r(s_i)+\sqrt{s_{i-1}}\,e^{-\sqrt{s_{i-1}}t}\,\r(s_{i-1})\right)\ ,
\nonumber
\end{eqnarray}
where $\r(s_1)=\r(s_{\rm th})=0$ and we used the trapezoidal rule for carrying out the
sums.

The maximum value $s_N$ is (approximately) equal to $m_\t^2$.\footnote{For the data set employed in Sec.~\ref{data} and Sec.~\ref{comp}
$s_N$ is slightly smaller than $m_\t^2$.}   For values of $s>s_N$ we use five-loop 
masslesss PT  \cite{PT} with an estimate for the order-$\a_s^5$ coefficient:
\begin{eqnarray}
\label{pertpart}
\bP_{\rm PT}(Q^2)&=&Q^2\int_{s_N}^\infty ds\,\frac{\r_{\rm PT}(s)}{s(s+Q^2)}\ ,\\
C_{\rm PT}(t)&=&\half\int_{s_N}^\infty ds\,\sqrt{s}\,e^{-\sqrt{s}t}\,\r_{\rm PT}(s)\ ,
\nonumber
\end{eqnarray}
where
\begin{equation}
\label{rhoPT}
\r_{\rm PT}(s)=\frac{1}{4\p^2}\left(1+\sum_{n=1}^5\sum_{m=1}^n c_{nm}J_m 
a^n(s;m_\t^2)\right)
\end{equation}
with $a(s;m_\t^2)\equiv\a_s(s;m_\t^2)/\p$ the strong coupling evolved from
$m_\t^2$ to $s$ using five-loop running \cite{5loop,5loop2}, 
\begin{equation}
\label{Js}
J_m=\p^{m-1}\sin(\p m/2)\ ,
\end{equation}
and, for QCD with three flavors,
\begin{equation}
\label{cs}
c_{11}=1\ ,\quad c_{21}=1.63982\ ,\quad c_{31}=6.37101\ ,\quad c_{41}=49.0757\ ,
\quad c_{51}=283\pm 141\ ,
\end{equation}
where the values for $c_{nm}$ for $m>1$, which follow from the renormalization 
group,\footnote{See for example Ref.~\cite{MJ}.} are given in
App.~\ref{rhoder}.
The value of $c_{51}$ has not been calculated, and instead, as indicated in 
Eq.~(\ref{cs}) we use an estimate \cite{BJ,BMO}.\footnote{For completeness, we include 
a brief review of the derivation of Eq.~(\ref{rhoPT}) in App.~\ref{rhoder}.} 
The values of the
$c_{mn}$ are those for $N_f=3$ perturbation theory.    While this is appropriate for
our comparison with $N_f=2+1$ flavor lattice results from Ref.~\cite{Mainz25} in 
Sec.~\ref{vacpol}, many of the lattice results we compare with in Sec.~\ref{windows}
have been obtained with $N_f=2+1+1$ flavor lattice QCD.   However, the effects of 
including or excluding dynamical charm effects appear to be very small 
\cite{Mainz22,Mainz24}, and we will ignore them in this paper.

At the $\t$ mass, quark-hadron duality violations (DV) are still visible
in the experimental spectral distribution, and we
thus introduce a DV component to the representation of the spectral function 
above $s=s_N$ to parametrize this
non-perturbative effect.   We use the model we have employed starting in 
Ref.~\cite{alphas1},
based on the theoretical considerations about the asymptotic behavior of
the spectral function of Ref.~\cite{BCGMP}.\footnote{Earlier work can be found in
Refs.~\cite{PQW,CGP,russians,russians2,russians3,catalans,CGP05,CGPmodel}.}
The explicit form is
\begin{equation}
\label{rhoDV}
\r_{\rm DV}(s)=e^{-\d-\g s}\,\sin(\a+\b s)\ ,
\end{equation}
where $\a$, $\b$, $\g$ and $\d$ are the ``DV'' parameters.
Both the perturbative part and the DV model will only
be employed above $s_N\approx m_\t^2$.   The expressions for $\bP_{\rm DV}(Q^2)$
and $C_{\rm DV}(t)$ simply follow by replacing $\r_{\rm PT}(s)$ by $\r_{\rm DV}(s)$
in Eq.~(\ref{pertpart}).

\subsection{\label{window} Light-quark-connected window quantities}

In addition to comparing the vacuum polarization $\bP(Q^2)$ obtained using $\t$ data 
with 
lattice results, it is also interesting to compare the lqc part of the HVP contribution 
$a_\m^{\rm HVP}$ to the 
anomalous muon magnetic moment and related window quantities \cite{RBC18}
obtained using the $\t$ data with results obtained on the lattice in isoQCD.

The lqc contribution to $a_\m^{\rm HVP}$ is given by
\begin{equation}
\label{amuHVP}
a_\m^{\rm HVP,lqc}=\frac{10}{9}\int_0^\infty dt\,w(t)C(t)\ ,
\end{equation}
where $C(t)$ is defined in Eq.~(\ref{Ct}), and $w(t)$ is the Bernecker-Meyer
kernel \cite{BM}, for which we use the sufficiently precise approximation of 
Ref.~\cite{Mainz17},
normalized such that for small $t$
\begin{equation}
\label{wt}
w(t)=\frac{1}{18}\, \a_{\rm EM}^2m_\m^2\left(t^4+\co(m_\m^2t^2)\right)\ .
\end{equation}
Window quantities are defined by inserting ``window'' functions $W(t)$ into 
Eq.~(\ref{amuHVP}),
\begin{equation}
\label{amuW}
a_\m^{\rm W,lqc}=\frac{10}{9}\int_0^\infty dt\,W(t)w(t)C(t)\ .
\end{equation}
Using 
\begin{equation}
\label{step}
\theta(t;t',\D)=\half\left(1+\tanh\left(\frac{t-t'}{\D}\right)\right)\ ,
\end{equation}
Ref.~\cite{RBC18} defined the short-distance (SD), intermediate (W1) and 
long-distance (LD) window functions as
\begin{eqnarray}
\label{wdefs}
W_{\rm SD}&=&1-\theta(t;t_0,\D)\ ,\\
W_{\rm W1}&=&\theta(t;t_0,\D)-\theta(t;t_1,\D)\ ,\nonumber\\
W_{\rm LD}&=&\theta(t;t_1,\D)\ ,\nonumber
\end{eqnarray}
choosing
\begin{equation}
\label{wpars}
t_0=0.4\ \mbox{fm}\ ,\quad t_1=1\ \mbox{fm}\ ,\quad \D=0.15\ \mbox{fm}\ .
\end{equation}
In Sec.~\ref{comp} we will compare results obtained using the $\t$-based,
IB-corrected spectral functions with lqc lattice results for $a_\m^{\rm HVP,lqc}$
and $a_\m^{\rm W,lqc}$, with W~=~SD, W1 and LD.

\subsection{\label{isospin} Isospin breaking}

In order to compare with lattice results for isospin-symmetric lqc quantities, 
we need to adjust the $\t$-data-based parts of $\bP(Q^2)$ and $C(t)$, defined in  
Eq.~(\ref{datapart}),
for IB.   The theoretical part above $s=s_N$ is already in pure isospin-symmetric
QCD,\footnote{Above $s=s_N$ the up and down quark masses can be neglected.}
but the data-based part needs to be corrected.    A first observation is that, working
to first order in IB, IB corrections in the $I=1$ vacuum polarization are purely 
electro-magnetic
(EM) in origin.  This follows because both currents in the lqc diagrams conserve
chirality, and thus two mass insertions of $m_u$ or $m_d$ are needed for a 
non-vanishing IB contribution.   If the insertions are on the same quark line,
the insertion will be proportional to $m_u^2+m_d^2$; if they are on different
quark lines, they are proportional to $m_um_d$.   These combinations can
be rewritten as $(m_u+m_d)^2\pm (m_u-m_d)^2$, and thus are second order
in strong IB. Another way to see this is to observe that the two 
$I=1$ currents in the correlator are $G$-parity positive, while the
SIB mass operator is $G$-parity negative. A contribution with
a single SIB mass insertion thus vanishes identically.    We will not explicitly use this observation below, but, for example,
it excludes the need to consider the form factor $f_-(s)$ that appears in the
differential hadronic $\t$ decay rate \cite{CEN} if we work to first order in IB.

For the case of the EM current correlator, IB corrections have been computed or
estimated on 
the lattice \cite{BMW20,Mainz24,ExtendedTwistedMass:2022jpw,RBC:2023pvn,
MILC:2024ryz}.
However, the breakdown into individual components needed to adjust these IB
corrections to the case of the $\t$-based representation is not available.   We thus
need to compare with lattice results in isoQCD, and adjust the $\t$-based
representation for IB effects in the $\t$-decay data.   Here we describe how we
estimate these effects, including a discussion of our estimate of the associated 
systematic errors.

We will separate out the two-pion component of the $I=1$ spectral function, 
$\r_V(s)\equiv\r_V^{I=1}(s)$, and, with the four-pion and other sub-dominant 
contributions expected to have only 
small IB corrections
of order $1\%$ or less, we will assume that IB in the remaining part of the 
spectrum can be neglected for the quantities considered in this paper.   Our task
is thus to correct the exclusive two-pion-mode contribution to $\rho_V(s)$
obtained from hadronic
$\t$ decays for IB.   For this, we first need to define
what is meant by pure isospin-symmetric QCD (isoQCD), also at low energies.  
We will
take the definition from Ref.~\cite{FLAG}, which for our purposes is equivalent to 
taking the ``WP25'' scheme of Ref.~\cite{WP25}.   In this scheme, we 
set the charged and neutral pion masses to $m_{\p^{\rm iso}}=m_{\p^0}=135$~MeV, and 
the charged and neutral kaon masses to $m_{K^{\rm iso}}=494.6$~MeV.\footnote{Kaon 
loop contributions
barely play a role in the IB corrections to the two-pion spectral function.}

An advantage of the current $\t$-based approach when it comes to estimating
IB corrections is the absence of the mixed-isospin, isoscalar-isovector 
contributions (such as those produced by $\r$-$\o$ mixing) 
present in measured $e^+ e^-\to \mbox{hadrons}(\g)$ cross sections. This makes 
the task of estimating IB corrections in the current
analysis somewhat simpler than that needed to estimate the IB corrections
required to convert the $\t$-based two-pion distribution to that measured in
$e^+ e^-\to\mbox{hadrons}(\g)$. 
We will base our approach on the recent study of Ref.~\cite{CCHH}, starting from the 
expression for the two-pion contribution to the spectral function
\begin{equation}
\label{2pisf}
\r_{[\t^-\to\p^-\p^0\n_\t(\g)]}(s)=S^{\p\p}_{\rm EW}\b_{\p^-\p^0}^3(s)|f_+(s)|^2 
G_{\rm EM}(s)\ ,
\end{equation}
where $S^{\p\p}_{\rm EW}$ accounts for short-distance electroweak corrections and
$G_{\rm EM}(s)$ accounts for radiative corrections in the decay.   The factor
$\b_{\p^-\p^0}^3(s)$ is the usual phase-space factor
\begin{eqnarray}
\label{phase}
\b_{\p^-\p^0}(s)&=&\l^{1/2}\left(1,\frac{m_{\p^-}^2}{s},\frac{m_{\p^0}^2}{s}\right)\ ,\\
\l(x,y,z)&=&x^2+y^2+z^2-2(xy+yz+zx)\ ,
\nonumber
\end{eqnarray}
and, finally, $f_+(s)$ is the weak pion form factor.   To first order in IB (\ie, 
first order in fine-structure constant $\a_{\rm EM}$ and the down-up quark mass 
difference $m_d-m_u$), Eq.~(\ref{2pisf}) is sufficient.  In the spectral 
functions to be used below \cite{us25}, the inclusive factor $S_{\rm EW}$ 
\cite{erlersew} has already
been removed.   Therefore, for the two-pion contribution, the relative additional
electroweak correction factor is equal to $S^{\p\p}_{\rm EW}/S_{\rm EW}$ 
(see Ref.~\cite{CHV} and references therein).

For the $s$-dependent EM correction $G_{\rm EM}(s)$ we employ the
dispersive determination of Ref.~\cite{CCHH}.
We interpolate the tabulated results\footnote{This table can be found at
{\tt https://arxiv.org/src/2511.07507v2/anc}.} using a cubic spline, as shown in 
Fig.~\ref{fig:GEM}.
Since $G_{\rm EM}(s)=1+\co(\a_{\rm EM})$, we can neglect the dependence on 
light quark or pion mass differences in $G_{\rm EM}-1$, as such effects will be second order in IB.

\begin{figure}[t!]
\vspace*{4ex}
\begin{center}
\includegraphics*[width=10.25cm]{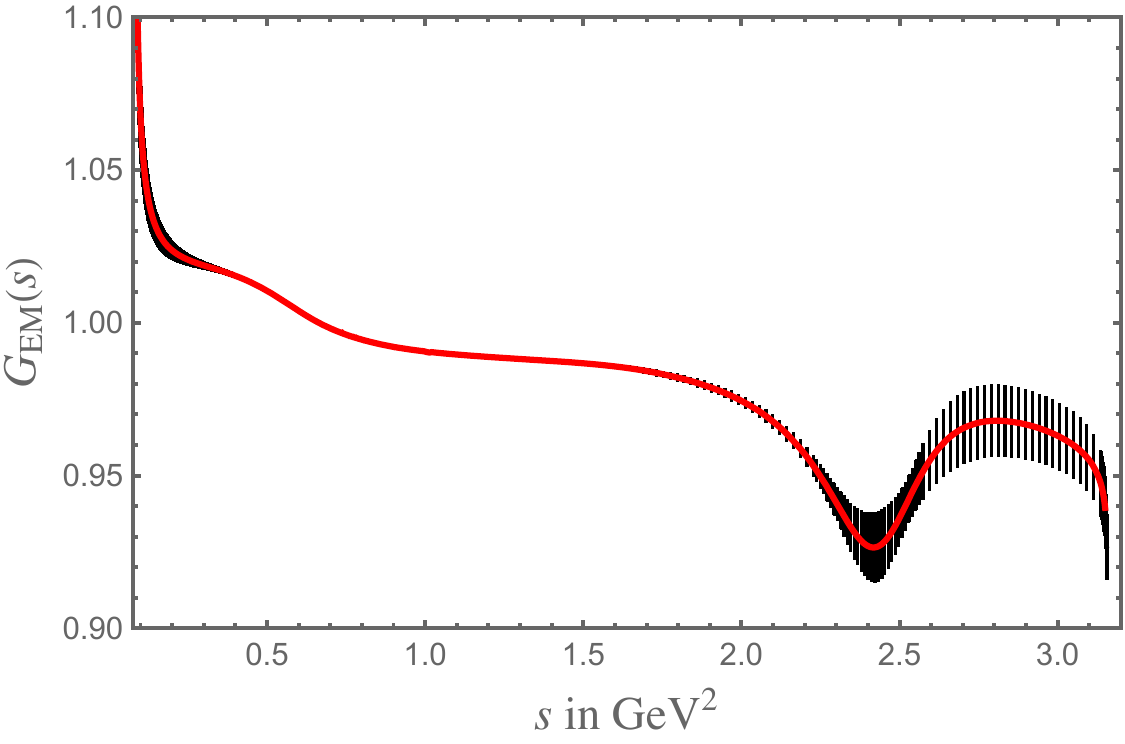}
\end{center}
\begin{quotation}
\floatcaption{fig:GEM}%
{{\it The radiative correction factor $G_{\rm EM}(s)$.  The black points are the 
tabulated values provided in Ref.~\cite{CCHH}; the red curve is a cubic spline 
interpolation of these points.}}
\end{quotation}
\vspace*{-4ex}
\end{figure}

The phase-space factor $\b_{\p^-\p^0}^3(s)$ in Eq.~(\ref{2pisf}) depends on the 
pion masses, and we thus include in our correction of $\r_{[\t^-\to\p^-\p^0\n_\t(\g)]}(s)$
a factor
\begin{equation}
\label{phasespcorr}
\frac{\b_{\p^0\p^0}^3(s)}{\b_{\p^-\p^0}^3(s)}\ ,
\end{equation}
in order to adjust the spectral function to that corresponding to isoQCD.

Similarly, the two-pion form-factor $f_+(s)$ depends on the pion masses.
Such an IB effect was not considered in Ref.~\cite{CCHH}, which was focused on a dispersive 
estimate of the radiative corrections, leaving IB contributions to $f_+(s)$
to future work. Here we investigate this
additional source of IB using the chiral-perturbation-theory (ChPT)-inspired
model of Ref.~\cite{CEN,Guerrero:1997ku} for which
\begin{equation}
\label{FF}
f^{\rm model}_+(s)=\frac{m_\r^2}{m_\r^2-s-im_\r\G_\r(s)}\,
\mbox{exp}[2\tilde{H}_{\p^-\p^0}(s)+\tilde{H}_{K^-K^0}(s)]\ ,
\end{equation}
in which
\begin{equation}
\label{Gammarho}
\G_\r(s)=\frac{m_\r s}{96\p f_\p^2}\left(\b_{\p^-\p^0}^3(s)\theta(s-(m_{\p^+}+
m_{\p^0})^2)+\half\b_{K^-K^0}^3(s)\theta(s-(m_{K^-}+m_{K^0})^2)\right)\ ,
\end{equation}
and 
\begin{equation}
\label{Htilde}
\tilde{H}_{ij}(s)=\frac{1}{f_\p^2}\,\Re h_{ij}(s,m_\r)\ ,
\end{equation}
with $h_{ij}(s,\m)$, which encapsulates one-loop effects from ChPT, given 
in App.~\ref{formulas} for masses $m_i$ and $m_j$.

In order to correct the $\t$-data-based parts of Eq.~(\ref{datapart}) to the world of
isoQCD, we thus need to also multiply $\r_{[\t^-\to\p^-\p^0\n_\t(\g)]}(s)$ by the
ratio of the form factor~(\ref{FF}) evaluated with both pion masses equal to the neutral 
pion mass and both kaon masses equal to the kaon mass of isoQCD,
to that evaluated with ``real-world'' pion and kaon masses.
The complete correction factor is thus
\begin{equation}
\label{fullcorr}
F_{\rm IB}(s)=\frac{S_{\rm EW}}{S_{\rm EW}^{\p\p}}\,\frac{1}{G_{\rm EM}(s)}\,
\frac{\b_{\p^0\p^0}^3(s)}{\b_{\p^-\p^0}^3(s)}\,\frac{|f^{\rm model}_+(s)[m_{\p^-}\to 
m_{\p^0};\ m_{K^-,K^0}\to m_{K^{\rm iso}}]|^2}{|f^{\rm model}_+(s)
[\mbox{real\ world}]|^2}\ .
\end{equation}
The factor $S_{\rm EW}$ in the numerator accounts for the fact that all spectral
data employed in this paper have already been corrected by the factor $S_{\rm EW}$, 
so only the relative difference between $S_{\rm EW}$ and $S_{\rm EW}^{\p\p}$ still
needs to be accounted for.

\begin{figure}[t!]
\vspace*{4ex}
\begin{center}
\includegraphics*[width=10.25cm]{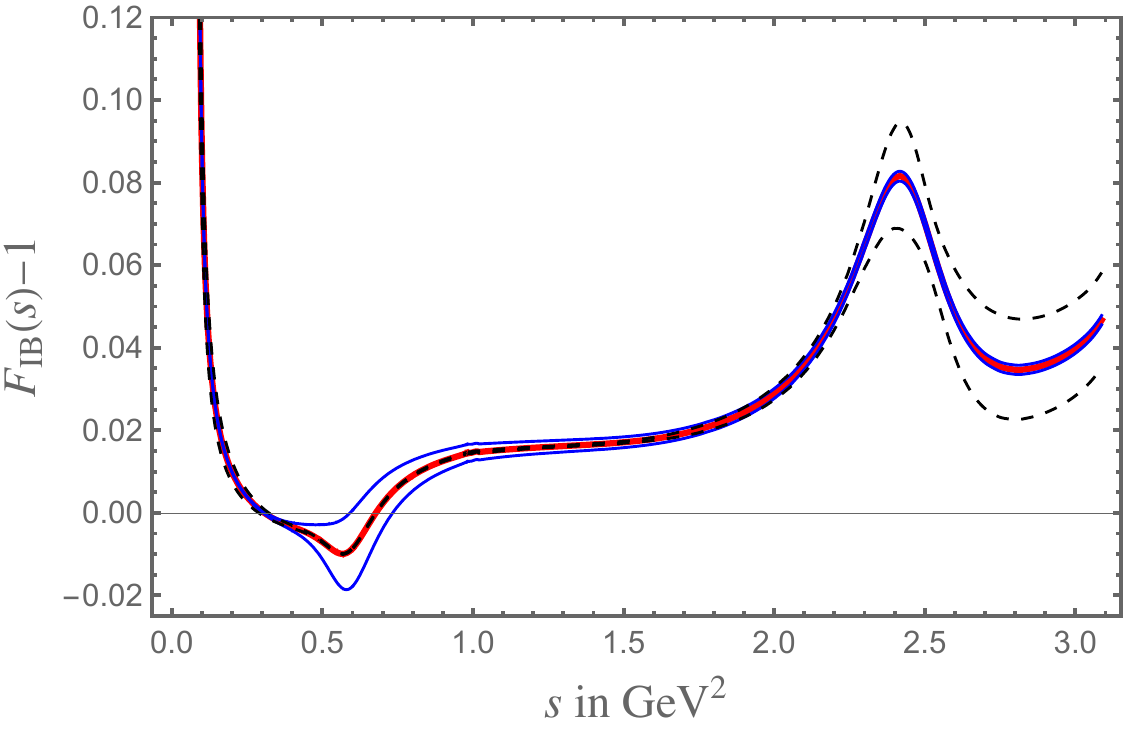}
\end{center}
\begin{quotation}
\floatcaption{fig:IB}%
{{\it The correction factor for isospin-breaking, Eq.~(\ref{fullcorr}), minus one.  The central value is 
shown in red, with the error band from Fig.~\ref{fig:GEM} in black (dashed thin curves) and the 
error band from varying $\G_\r$ in Eq.~(\ref{Gammarho}) by a factor $1\pm 2\a_{\rm eEM}/\p$ in blue (solid thin curves).}}
\end{quotation}
\vspace*{-4ex}
\end{figure}

Finally, the threshold $s_{\rm th}=4m_{\p^0}^2$ for $I=1$ hadronic $\t$ decays in isoQCD
is lower than the threshold $s_{\rm th}=(m_{\p^-}+m_{\p^0})^2$ in the real world, and
we need to correct for this.   For such small $s$, this correction can be reliably 
calculated using the phase space factor $\b^3_{\p^0\p^0}(s)$ and the form factor
of Eq.~(\ref{FF}) evaluated with both pion masses set equal to $m_{\p^0}$.   It turns out
that the thus-obtained isoQCD spectral function matches very well to the lowest
data points, after they have been corrected as described above.   It also turns out
that this threshold correction is so small that it can be safely neglected.

The use of the model of Ref.~\cite{CEN} for the form-factor $f_+(s)$ is the weakest
step in modeling IB corrections.  To explore the impact of possible changes in the form
used for the $s$-dependent width,
we have replaced the factor $m_\r s$ by
$m_\r^2\sqrt{s}$, following the form for $\G_\r$ used in Ref.~\cite{Davier10}; we find
that this has no significant numerical impact.   We have also followed the prescription 
of Ref.~\cite{WP25}
for modeling unknown EM corrections to the $\r\p\p$ coupling which affect $\r$ decays 
by multiplying
$\G_\r$ by the factor $1\pm 2\a_{\rm EM}/\p$.   This is done only in the 
denominator of Eq.~(\ref{fullcorr}), and not in the numerator, as the latter 
corresponds to the world of isoQCD in which there are no EM corrections and thus
also no uncertainties due to EM effects.
It turns out that this is a significant effect, as discussed in Ref.~\cite{WP25}, and we will 
use the size of this
effect as an estimate of the systematic error associated with using Eq.~(\ref{fullcorr})
in Sec.~\ref{comp} below. We show the IB correction factor, together wih the
errors from $G_{\rm EM}$ and from the prescription above, in Fig.~\ref{fig:IB}.
In addition, in order to account for other IB effects not 
covered in the discussion above (such as EM corrections to the $\r$ mass), we will 
add (in 
quadrature) an 
additional systematic error of $50\%$ of the total IB correction itself, on the
quantities to be computed in Sec.~\ref{comp}. We have 
checked, for example, that this covers the effect of varying the $\r$ mass independently
in the numerator and denominator of Eq.~(\ref{fullcorr}) within the range allowed by the
errors on the $\r^0$ and $\r^+$ masses given in Ref.~\cite{PDG}.

\section{\label{data} Data}

In this section, we will specify our data input.   We will be using the results from
Ref.~\cite{us25} for the vector $I=1$ exclusive-mode spectral function contributions, 
reviewed in more
detail in Sec.~\ref{decay} for the two-pion and inclusive cases, and in Sec.~\ref{fourpi}
for the four-pion case.  Here we summarize other input of our analysis.   

The main parameter values we use are
\begin{eqnarray}
\label{pars}
\a_s(m_\t^2)&=&0.317(6)\ ,\\
m_\t&=&1.77693~\mbox{GeV}\ ,\nonumber\\
\a_{\rm EM}&=&\frac{1}{137.036}\ ,\nonumber\\
m_\m&=&0.105658~\mbox{GeV}\ .
\nonumber
\end{eqnarray}
Here $\a_s(m_\t^2)$ is taken from Ref.~\cite{FLAG6}, evolved to the $\t$ mass
and quantities shown without errors are so precisely known that their errors
do not affect our analysis.

For IB corrections, we take the additional parameter values to be
\begin{eqnarray}
\label{IBpars}
m_{\p^-}&=&0.13957039~\mbox{GeV}\ ,\\
m_{\p^0}&=&0.1349768~\mbox{GeV}\ ,\nonumber\\
m_{K^-}&=&0.493677~\mbox{GeV}\ ,\nonumber\\
m_{K^0}&=&0.497611~\mbox{GeV}\ ,\nonumber\\
m_{K^{\rm iso}}&=&0.4946~\mbox{GeV}\ ,\nonumber\\
m_\r&=&0.775~\mbox{GeV}\ ,\nonumber\\
f_\p&=&0.09221~\mbox{GeV}\ ,\nonumber\\
S_{\rm EW}&=&1.0201\ ,\nonumber\\
S_{\rm EW}^{\p\p}&=&1.0233\ ,
\nonumber
\end{eqnarray}
where all values come from Ref.~\cite{PDG} except $m_{K^{\rm iso}}$ which 
comes from Ref.~\cite{FLAG6}, $S_{\rm EW}$ which is taken from Ref.~\cite{erlersew}, and
$S_{\rm EW}^{\p\p}$, for which we  follow the discussion of  Ref.~\cite{CHV}. Strictly 
speaking the pion mass in isoQCD
is defined to be $0.135~$GeV in Ref.~\cite{FLAG6}, but the difference with
$m_{\p^0}$ given above is too small to have any effect.  Errors on all these 
values are negligibly small for our purposes.

We will also need values for the parameters in Eq.~(\ref{rhoDV}).   We take these
from a fit of the $w=1$ moment \cite{us25} at $s_0^{\rm min}=1.5747$~GeV$^2$ 
fixing $\a_s(m_\t^2)=0.317$:
\begin{eqnarray}
\label{DVpars}
\d&=&3.3857\ ,\\
\g&=&0.62575~\mbox{GeV}^{-2}\ ,\nonumber\\
\a&=&-0.20412\ ,\nonumber\\
\b&=&3.2111~\mbox{GeV}^{-2}\ .\nonumber
\end{eqnarray}
The uncertainties on these values are not small but they are not very important, 
as we will take the difference of
the effect of including DVs with these parameters and omitting DVs completely
as the systematic error on the DV contribution.

\subsection{\label{decay} Hadronic \begin{boldmath}$\t$-decay\end{boldmath} data}

Our hadronic spectral data for the  $I=1$ vector spectral function are obtained by
combining data from a range of experiments.   First, for the $\t^-\to\p^-\p^0\n_\t$ 
spectral
distribution, we combine data from ALEPH \cite{ALEPH,ALEPH2,ALEPH13}, 
OPAL \cite{OPAL} and Belle \cite{Belle}, as
described in detail in Ref.~\cite{us25}.\footnote{The inclusion of CLEO data \cite{CLEO,
CLEO2pi} was investigated in an appendix of Ref.~\cite{us25}, neglecting the unknown 
systematic correlations, and the impact of CLEO data was found to be small.}
  We use the $2\p$ spectral distribution to estimate IB 
corrections, as described in Sec.~\ref{isospin}.   Then, ALEPH and OPAL data for
the decays $\t^-\to2\p^-\p^+\p^0\n_\t$ and $\t^-\to\p^-3\p^0\n_\t$ are combined
into a single $4\p$ spectral function; this is again described in detail in Ref.~\cite{us25}.
In Sec.~\ref{fourpi} below we will compare the mode-by-mode ALEPH and OPAL four-pion 
spectral distributions with those  implied by four-pion
electroproduction data,  which are related by isospin
symmetry as expressed through the Pais relations ({\it cf.} Eq.~(\ref{Paisrel})).   

Contributions from the remaining (``residual'') modes have also all been determined from 
experimental data, as described in Ref.~\cite{alphas20}; none of the contributions from these
modes is obtained from Monte-Carlo simulation. This represents an improvement 
over
the ALEPH and OPAL treatments of 
contributions from these modes, where, lacking such experimental input, 
Monte Carlo estimates had to be 
employed.  The sum of the two-pion, four-pion and residual-mode spectral distributions 
obtained in Ref.~\cite{us25} yields the inclusive spectral 
function to be used in the data parts of Eqs.~(\ref{vacpolQ2}) and~(\ref{Ct}).

\begin{figure}[t!]
\vspace*{4ex}
\begin{center}
\includegraphics*[width=10.25cm]{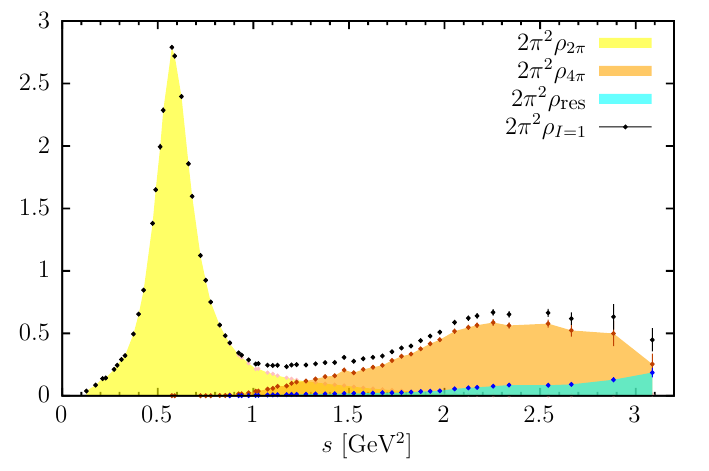}
\end{center}
\begin{quotation}
\floatcaption{fig:incl2pi}%
{{\it Rescaled inclusive $I=1$ spectral function $2\p^2\r(s)$, the two-pion 
mode 
$2\p^2\r_{2\p}(s)$ (red area), the four-pion modes $2\p^2\r_{4\p}(s)$ (green area),
and the residual modes $2\p^2\r_{\rm res}(s)$ (purple area)
from vector hadronic $\t$ decays,
from Ref.~\cite{us25}.   The normalization shown is such that the parton-model value of
the rescaled inclusive spectral function is  $1/2$.}}
\end{quotation}
\vspace*{-4ex}
\end{figure}

We show the inclusive spectral function as well as the two-pion, combined 
four-pion and combined residual spectral distributions in Fig.~\ref{fig:incl2pi}.
The four-pion exclusive-mode spectral function
contributions will be shown in the next subsection.

\subsection{\label{fourpi} Spectral distributions for the four-pion modes}

With the recent focus on the contribution from the two-pion mode to $a_\m^{\rm HVP}$
\cite{WP25},
discrepancies between the spectral distributions for $\t^-\to\p^-\p^0\n_\t$ and 
$e^+e^-\to\p^+\p^-$ have been studied extensively (see for example 
Ref.~\cite{Davieretal23}).  Here we consider the two four-pion modes
in more detail, comparing contributions to $\rho_V(s)$ obtained from the 
$\t^-\to\p^-3\p^0\n_\t$ and $\t^-\to 2\p^-\p^+\p^0\n_\t$ decay distributions 
to those implied by the $I=1$, EM spectral function contributions obtained 
from measured $e^+e^-\to 2\p^+2\p^-$ and $e^+e^-\to\p^+\p^-2\p^0$ 
cross sections. With exact isospin symmetry, the EM expectations for
the $\tau$ exclusive-mode contributions are given by the Pais 
relations
\cite{Pais,Aleph4pi}
\begin{subequations}
\begin{eqnarray}
\label{Paisrel}
&&\hspace{-1.8cm}\r_V[\t,2\p^-\p^+\p^0](s)=\half\,\r_V[\mbox{EM},2\p^-2\p^+](s)+
\r_V[\mbox{EM},\p^-\p^+2\p^0](s)\ ,\label{Paisrela}\\
&&\hspace{-1.8cm}\r_V[\t,\p^-3\p^0](s)=\half\,\r_V[\mbox{EM},2\p^-2\p^+](s)
\ .\label{Paisrelb}
\end{eqnarray}
\end{subequations}
The quantities on the right-hand side of these equations
differ by a factor 2 from the usual EM four-pion spectral function contributions, as the
$I=1$ part of the EM current is $1/\sqrt{2}$ times the neutral-current partner of the
charged vector $ud$ current.

It is unknown how to systematically correct these relations for IB.   However,
since the support of the four-pion spectral distributions has little overlap with the 
$\r-\o$ narrow-resonance region, one expects these relations to hold to better
than $1\%$ (the typical size of IB corrections) also when IB is taken into account. 

\begin{figure}[t!]
\vspace*{4ex}
\begin{center}
\includegraphics*[width=7.3cm]{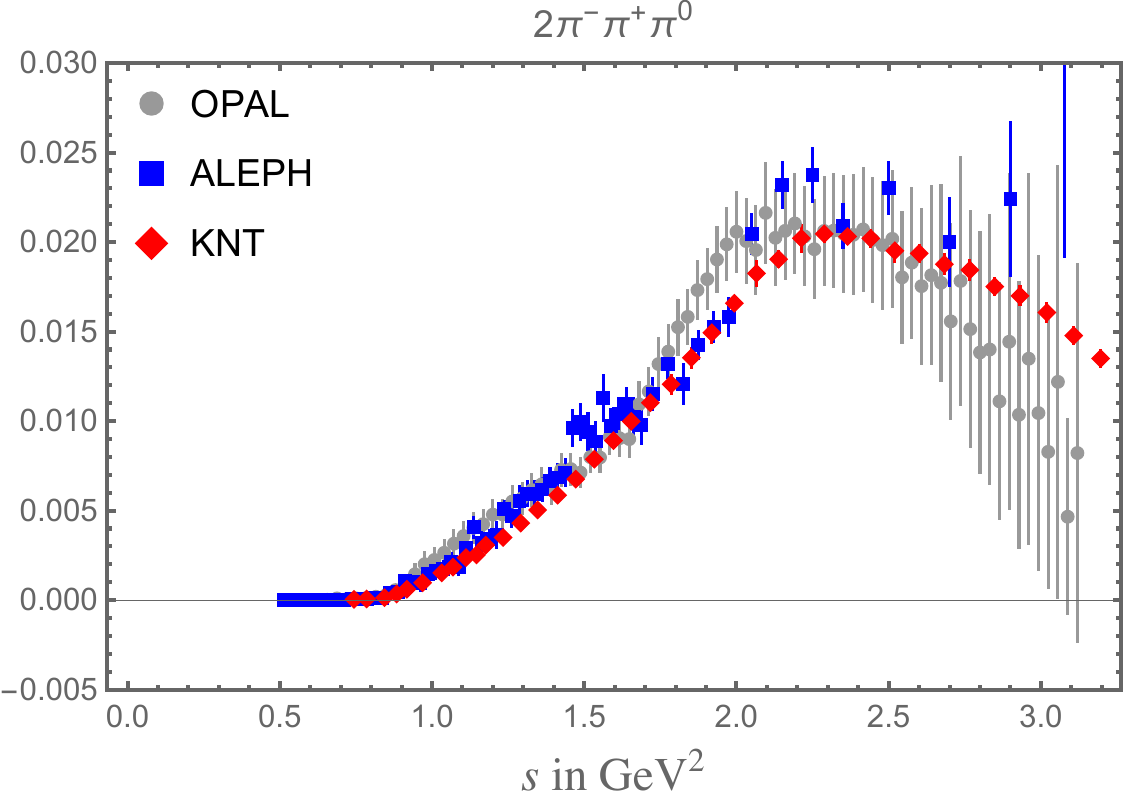}
\hskip0.5cm
\includegraphics*[width=7.3cm]{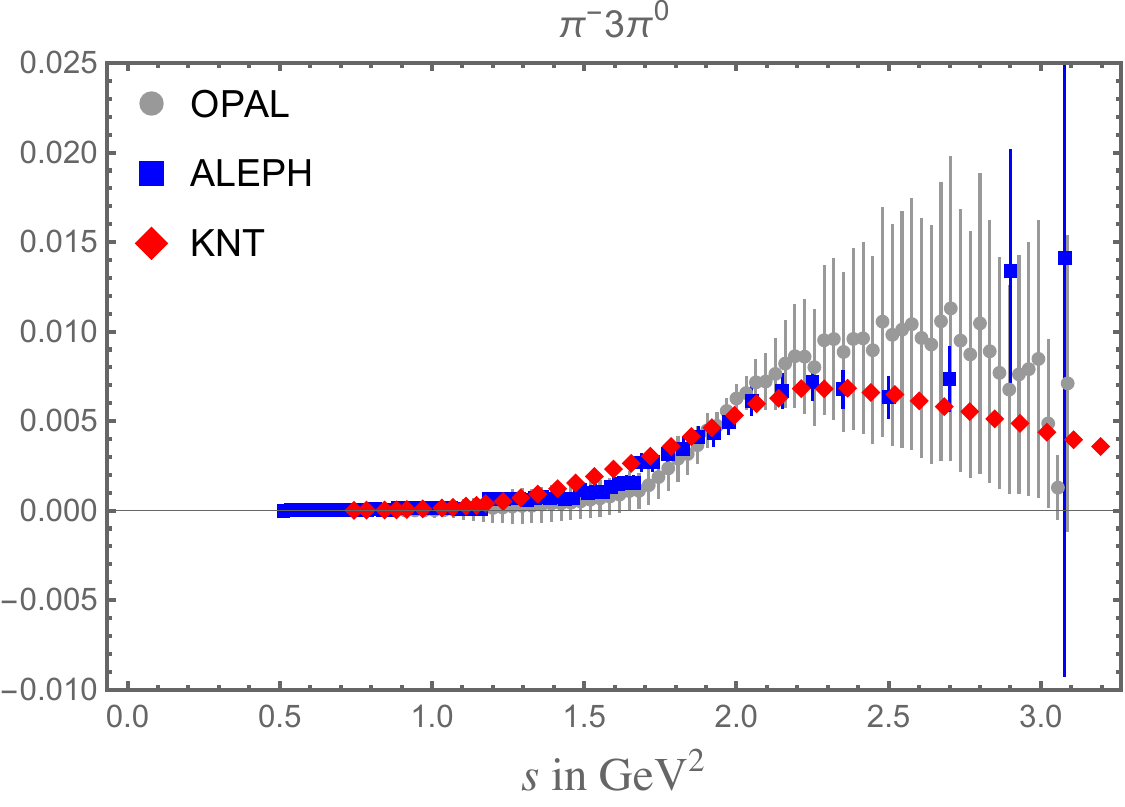}
\end{center}
\begin{quotation}
\floatcaption{fig:4pialephopal}%
{{\it $2\p^-\p^+\p^0$ (left panel) and $\p^-3\p^0$ (right panel) ALEPH (blue) and
OPAL (gray) $\t$-based spectral functions, compared with the $e^+e^-$-based
combinations on the right-hand sides of Eqs.~(\ref{Paisrela})
and~(\ref{Paisrelb}), obtained using KNT19 data (red).}}
\end{quotation}
\vspace*{-4ex}
\end{figure}

\begin{figure}[t!]
\vspace*{4ex}
\begin{center}
\includegraphics*[width=7.3cm]{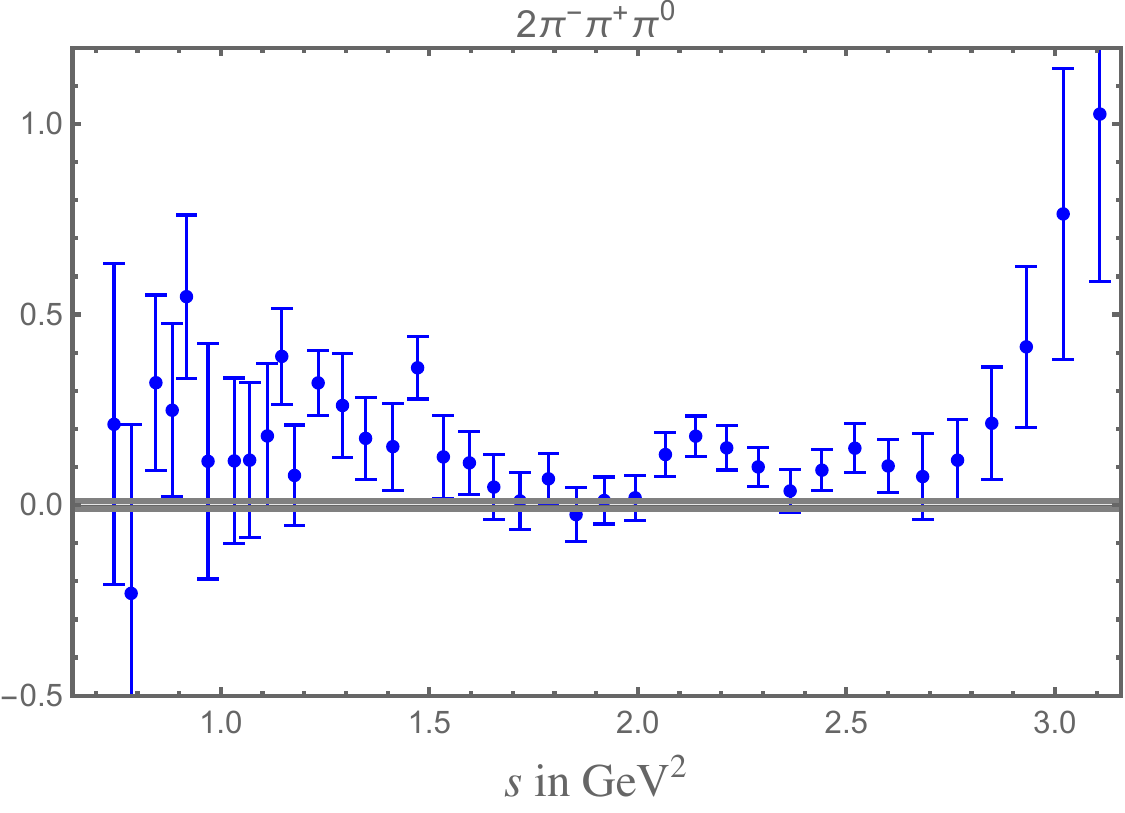}
\hskip0.5cm
\includegraphics*[width=7.3cm]{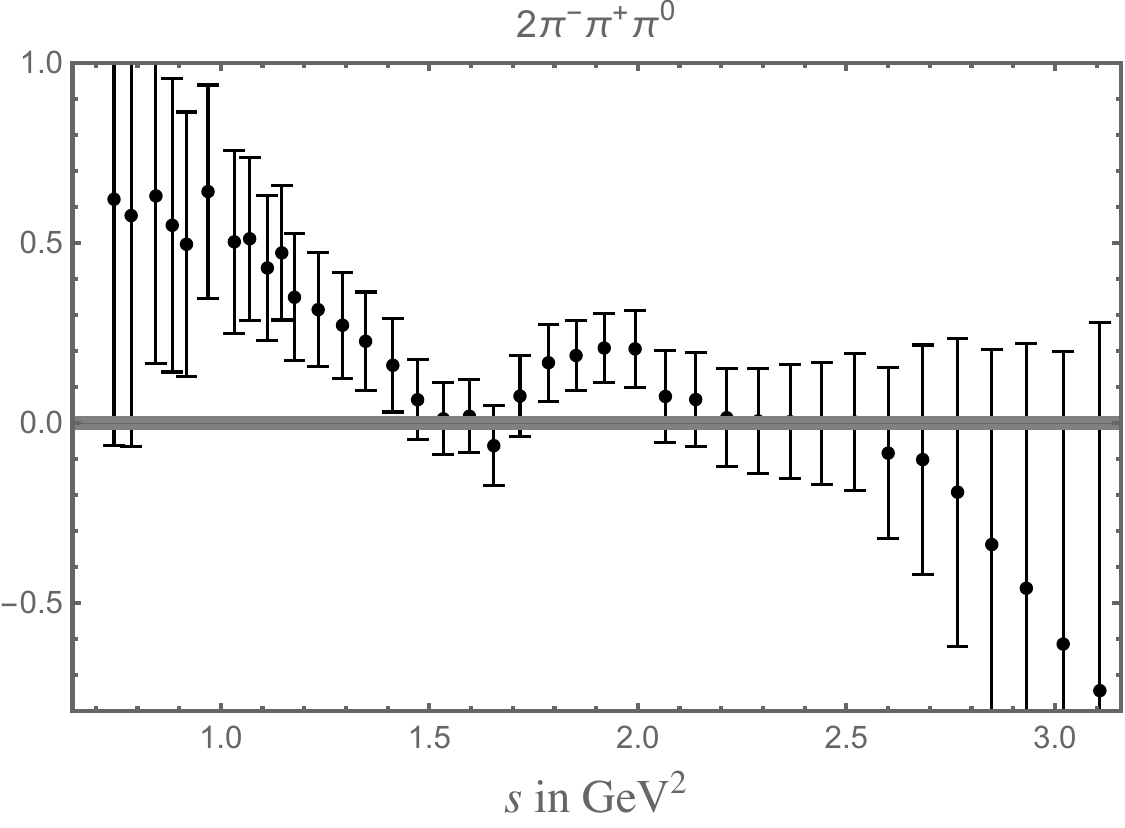}
\vskip0.5cm
\includegraphics*[width=7.3cm]{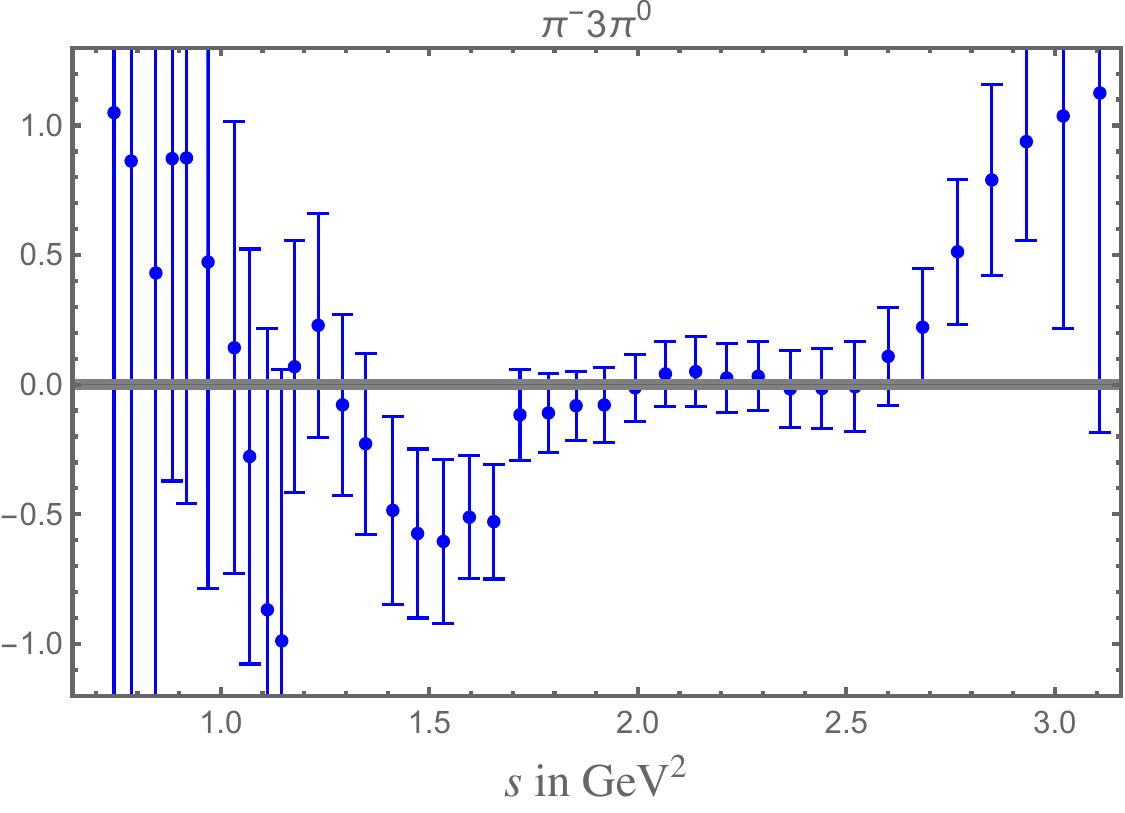}
\hskip0.5cm
\includegraphics*[width=7.3cm]{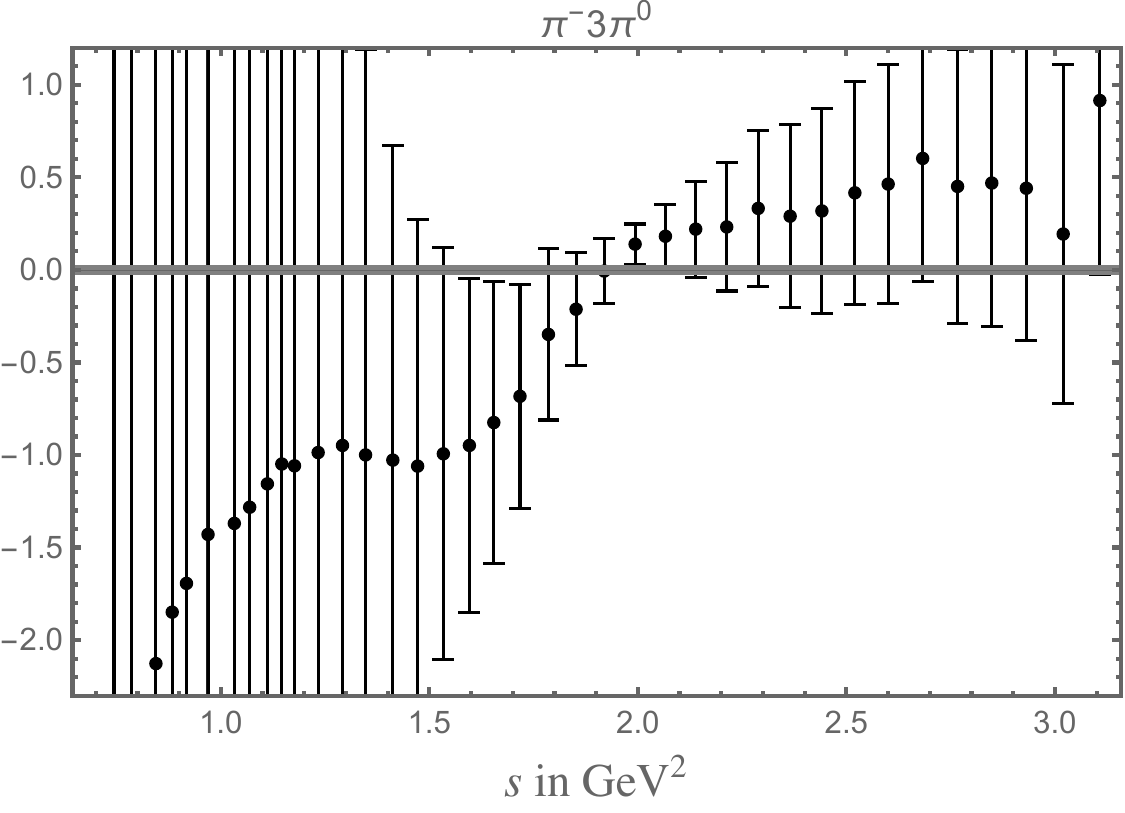}
\end{center}
\begin{quotation}
\floatcaption{fig:4pipull}%
{{\it The ratios $R^{\rm exp,mode}(s)$ defined in Eq.~(\ref{ratio}), for experiments
ALEPH (left) and OPAL (right) and modes $2\p^-\p^+\p^0$ (top) and
$\p^-3\p^0$ (bottom).   The gray horizontal band shows the region of 
values between $-0.01$ and $+0.01$.}}
\end{quotation}
\vspace*{-4ex}
\end{figure}

A comparison of the $\tau$-based ALEPH/OPAL four-pion results and
corresponding Pais relation expectations is shown in Fig.~\ref{fig:4pialephopal}.  
The Pais-relation expectations 
have been evaluated using the KNT19 four-pion exclusive-mode distribution results
provided to us by the authors of Ref.~\cite{KNT19}. The points labelled ``ALEPH'' 
and ``OPAL'' are the ALEPH and OPAL versions of the left-hand sides of 
Eq.~(\ref{Paisrel}) and those labelled ``KNT19'' the KNT19-based versions 
of the right-hand sides. In Fig.~\ref{fig:4pipull} we show the
relative differences between the ALEPH and OPAL four-pion spectral distributions 
and the KNT19-based Pais 
relation expectations after interpolation of the ALEPH and OPAL results to 
the $s$ values of the KNT19 set.   These relative differences are defined as follows.
Given a $\t$-based distribution $\r_\t^{\rm exp,mode}(s)$ and the corresponding
KNT19-based expectation, $\r_{e^+e^-}^{\rm mode}(s)$, where ``exp'' is 
ALEPH or OPAL, and
``mode" is $2\p^-\p^+\p^0$ or $\p^-3\p^0$, we define the ratios
\begin{equation}
\label{ratio}
R^{\rm exp,mode}(s)=\frac{\r_\t^{\rm exp,mode}(s)-
\r_{e^+e^-}^{\rm mode}(s)}{\half(\r_\t^{\rm exp,mode}(s)+\r_{e^+e^-}^{\rm mode}(s))}\ .
\end{equation}
If the difference between $\r_\t^{\rm exp,mode}(s)$ and $\r_{e^+e^-}^{\rm mode}(s)$
is explained by IB, we expect the corresponding ratio to lie between approximately
$-0.01$ and $+0.01$, shown in Fig.~\ref{fig:4pipull} by the gray horizontal bands
centered at the horizontal axis.

These figures show that there are significant discrepancies between the $\t$-based
data and $e^+e^-$ based data, in particular for the $2\p^-\p^+\p^0$ mode for 
which errors are smaller.  This may not be immediately obvious from Fig.~\ref{fig:4pipull},
in part because the figure does not show correlations between different points.  

In Tables~\ref{tab:windows1pi0} (for Eq.~(\ref{Paisrela})) and \ref{tab:windows3pi0} 
(for Eq.~(\ref{Paisrelb})) we show the results 
of the integrals with weights defined in Eq.~(\ref{amuW}) for the SD, W1 and LD 
windows as well
as the integral with $W=1$ in Eq.~(\ref{amuW}) for the four-pion spectral distributions 
from ALEPH
and OPAL, compared with the same integrals over the KNT19 EM combinations
on the right-hand side of these equations.   From these tables, we conclude that 
there are significant differences for the $2\p^-\p^+\p^0$ mode between the 
ALEPH and KNT19 data, ranging between $3.3\s$ for the SD window to $4.4\s$
for the LD window.   For the OPAL data, which has larger errors, the differences 
are less pronounced, 
varying between $0.4\s$ and $2.8\s$ for the SD and LD windows.  For the $\p^-3\p^0$ 
mode there is 
good agreement between the $\t$-based and $e^+e^-$-based window values.
We note, however, that the $\t$-based window values have much larger
relative errors for the $\p^-3\p^0$ mode than for the $2\p^-\p^+\p^0$ mode.
The results of Tables~\ref{tab:windows1pi0} and ~\ref{tab:windows3pi0} are
illustrated in Fig.~\ref{fig:Paistension}, where we plotted the $\t$-based values 
shown in the tables normalized to the KNT19 values.

\begin{table}[!t]
    \begin{center}
    \begin{tabular}{lcccc}
        \hline
        Data set & SD  & W1 & LD & HVP \\
        \hline
        ALEPH $1\pi^0$ & $5.37(19)$  & $17.06(44)$ & $5.157(92)$ & $27.58(70)$ \\
        OPAL $1\pi^0$  & $4.77(36)$  & $15.96(69)$ & $5.23(25)$  & $26.0(1.1)$  \\
        KNT19 $1\pi^0$   & $4.63(12)$  & $14.80(39)$ & $4.45(13)$  & $23.88(63)$ \\
        \hline
    \end{tabular}
    \end{center}
        \floatcaption{tab:windows1pi0}{{\it Integrals over the $\tau$ data for the 
        $2\pi^-\pi^+\pi^0$ spectral distribution and over the KNT19 EM spectral distributions in the combination shown on the right-hand side of Eq.~(\ref{Paisrela}) 
        with SD, W1, LD and HVP weights, in units of $10^{-10}$. The $\tau$ 
        data have been  linearly interpolated to KNT19 grid. The data are integrated from 
        $0.743906$~{\rm GeV}$^2$ to $3.01983$~{\rm GeV}$^2$, which avoids 
        extrapolation of the $\tau$ data.}
       \vspace{10mm}}
\end{table}

\begin{table}[!t]
    \begin{center}
    \begin{tabular}{lcccc}
        \hline
        Data set & SD  & W1 & LD & HVP \\
        \hline
        ALEPH $3\pi^0$ & $1.59(22)$   & $4.52(49)$   & $1.076(97)$  & $7.18(79)$  \\
        OPAL $3\pi^0$  & $1.61(56)$   & $4.5(1.0)$   & $0.98(14)$   & $7.1(1.5)$  \\
        KNT19 $3\pi^0$   & $1.361(18)$  & $4.164(56)$  & $1.104(16)$  & $6.629(90)$ \\
        \hline
    \end{tabular}
    \end{center}
        \floatcaption{tab:windows3pi0}{{\it Integrals over the $\tau$ data for the 
        $\pi^-3\pi^0$ spectral distribution and over the KNT19 EM spectral distribution 
        on the right-hand side of Eq.~(\ref{Paisrelb}) with SD, W1, LD and HVP weights, 
        in units of $10^{-10}$. The $\tau$ data have been  linearly interpolated to KNT19 
        grid. The data are integrated from $0.883468$~{\rm GeV}$^2$ to 
        $3.01983$~{\rm GeV}$^2$, which avoids extrapolation of the $\tau$ data.}}
\end{table}

\begin{figure}[t!]
\vspace*{4ex}
\begin{center}
\includegraphics*[width=8.0cm]{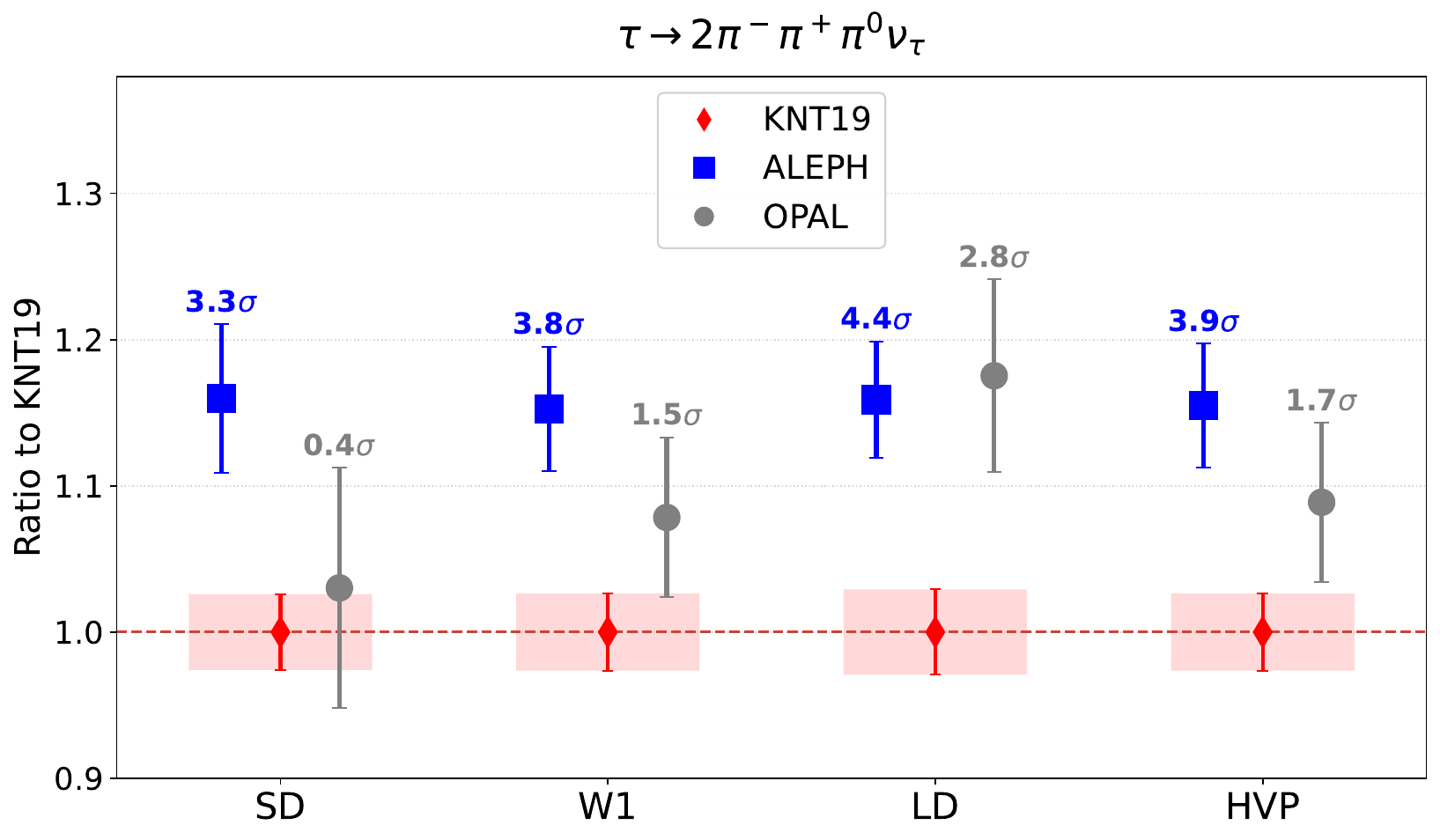}
\hskip0.3cm
\includegraphics*[width=8.0cm]{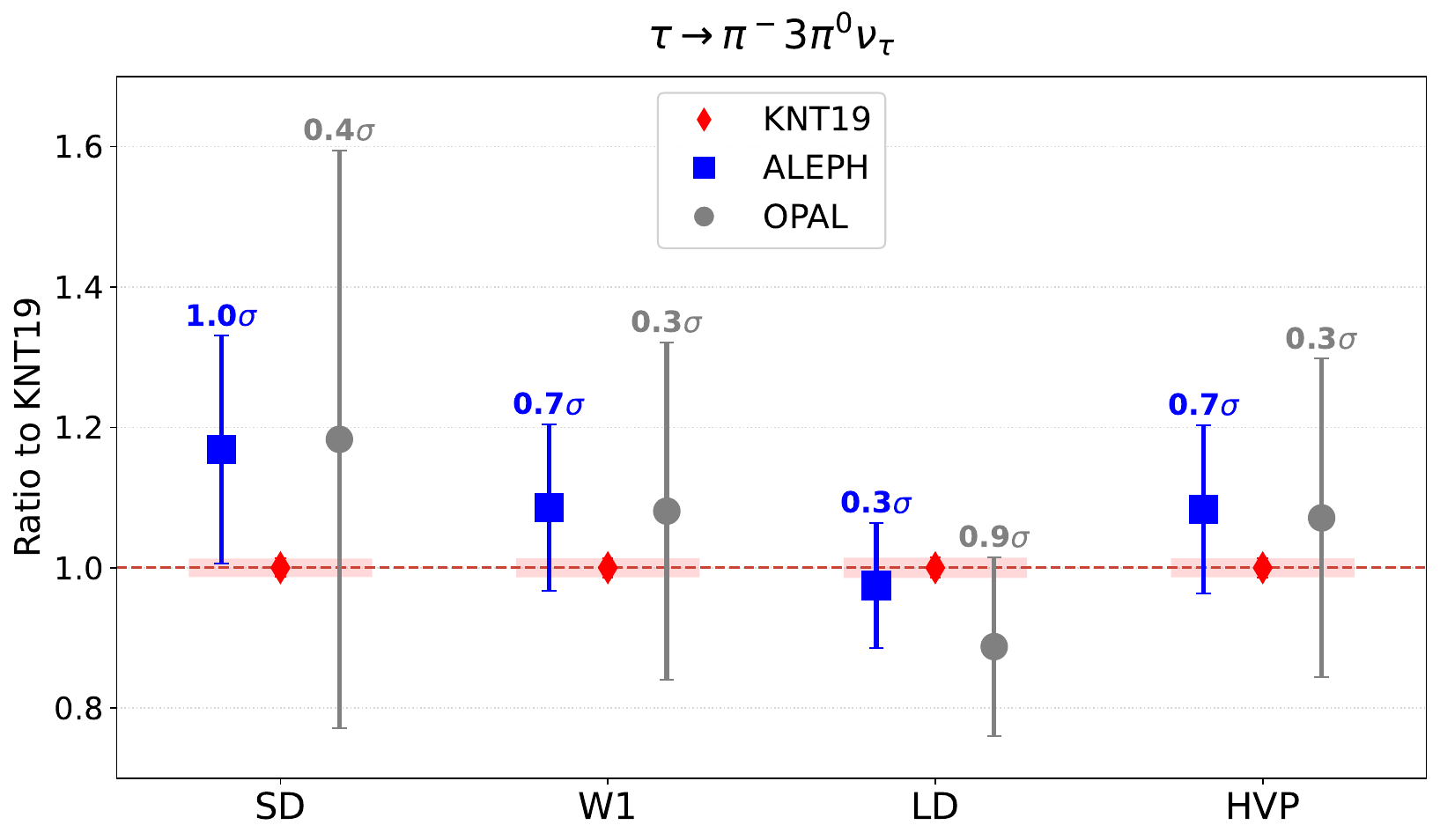}
\end{center}
\begin{quotation}
\floatcaption{fig:Paistension}%
{{\it Visual illustration of Tables \ref{tab:windows1pi0} and \ref{tab:windows3pi0}.
All results normalized to KNT19 central values.}}
\end{quotation}
\vspace*{-4ex}
\end{figure}

\section{\label{comp} Comparisons with lattice QCD}

In this section, we compare $\t$-based dispersive results with results 
from lattice QCD.   

Sections Sec.~\ref{vacpol} and Sec.~\ref{windows} contain the comparisons
for $\bP(Q^2)$, and the window quantities and $a_\m^{\rm HVP,lqc}$, 
respectively, with dispersive contributions from the $\tau$ kinematic 
region $s\le m_\tau^2$ evaluated using inclusive $\tau$ vector isovector 
data. The lattice results in Sec.~\ref{vacpol}
come primarily from the Mainz collaboration \cite{Mainz25}, with a few results 
also available from the BMW collaboration \cite{BMW20}.   Recently, preliminary 
work on 
$\bP(Q^2)$ was also reported by the Fermilab-MILC collaboration \cite{FNALMILC},
but no quantitative results are available yet for comparison.   The lattice results in
Sec.~\ref{windows} are taken from Ref.~\cite{WP25}.
In Sec.~\ref{exclusive} we perform alternate comparisons in which 
$\tau$ dispersive input to the data-based parts of the various dispersive 
integrals is restricted to either exclusive two-pion- or exclusive two-pion- 
and four-pion-mode contributions \cite{us25}. Contributions from all other 
exclusive modes, as well as from two-pion and four-pion contributions in the
region $s>m_\tau^2$, are obtained using KNT19 data \cite{KNT19} up to the 
KNT19 endpoint $s_{\rm KNT}$, with the perturbative form of 
the inclusive spectral function, supplemented by our estimate of possible 
DV corrections, used above that. In this subsection, we will also evaluate 
the four-pion contributions to the window quantities and $a_\m^{\rm HVP,lqc}$  
with $\t$-based input on the left-hand sides and KNT19 input on the right-hand 
sides. This will provide additional quantitative measures of the impact of the 
observed discrepancies between $\t$-based and $e^+e^-$-based measurements 
of the four-pion spectral distributions on quantities of interest in the 
determination of $a_\mu^{\rm HVP,lqc}$.

\subsection{\label{vacpol} Subtracted vacuum polarization 
\begin{boldmath}$\bP(Q^2)$\end{boldmath}}

In Table~\ref{tab:PiMainz} we compare results for $\bP(Q^2)$ from the 
$\t$-based dispersive representation~(\ref{vacpolQ2}) to those obtained on 
the lattice by the Mainz Collaboration \cite{Mainz25}, for values of $Q^2$ between 
$0.5$~GeV$^2$ and $12$~GeV$^2$.    The 
latter are both continuum extrapolated and finite-volume-corrected.

The errors on the data-based contributions to the dispersive results 
were obtained from the full covariance matrix of the input
inclusive spectral function.

Perturbative contributions were evaluated using the input $\a_s(m_\t^2)$ 
specified in Eq.~(\ref{pars}), with errors equal to the  
quadrature sum of (i) the error produced by propagating the 
uncertainty in Eq.~(\ref{pars}) through the computation, and (ii) a systematic error
equal to the maximum variation produced by either turning off the $\co(\a_s^5)$ term in
Eq.~(\ref{rhoPT}) altogether, or varying $c_{51}$ within the bounds specified in 
Eq.~(\ref{cs}).  The shift that results from varying $c_{51}$ is, for all
$Q^2$, largest when $c_{51}$ takes on its lowest allowed value in this range.

DV contributions were evaluated using the input parameters values 
shown in Eq.~(\ref{DVpars}). A 100\% systematic uncertainty, equal to the 
shift produced by turning off DVs altogether, is assigned to these 
estimates.   

There are no correlations between the data, PT and DV sources of error.

%%%%%%%%%
\begin{table}[t]
\begin{center}
\begin{tabular}{crcrc}
\hline
$Q^2$\ (GeV$^2$) & $\bP(Q^2)\qquad$  & IB corr  & $2\times\bP_{\rm M}(Q^2)$  &
pull \\
\hline
0.5 & 4490(20)(22)[30] & $-$29(22)  & 4634(42) & 2.8 \\
1  & 6366(29)(31)[42] & $-$39(31) & 6534(38)& 3.0 \\
2  & 8345(40)(37)[55] & $-$48(37)  & 8528(46) & 2.6 \\
3  & 9503(48)(40)[63] & $-$51(40)  & 9688(52)&  2.3 \\
4  & 10318(53)(42)[68] & $-$53(42)  & 10530(40) & 2.7 \\
5  & 10946(57)(43)[72] & $-$54(43)  & 11162(40)& 2.6  \\
6  & 11456(61)(44)[75] & $-$55(44)   & 11674(42) & 2.5 \\
7  & 11886(63)(44)[77] & $-$56(44)  & 12102(44)& 2.4  \\
8  & 12257(65)(45)[79] & $-$56(45) & 12470(46)& 2.3 \\
9  & 12583(67)(45)[81] & $-$57(45)  & 12796(48) & 2.3 \\
12  & 13377(71)(46)[84] & $-$58(46)  & 13586(52)& 2.1 \\
\hline
\end{tabular}
\end{center}
\floatcaption{tab:PiMainz}%
{{\it Comparison of $\t$-based $\bP(Q^2)$ values (which include the IB correction 
shown in the 3rd column) with the lattice values 
corresponding to the $\bP_{\rm M}(Q^2)$ results from Table~5 of 
Ref.~\cite{Mainz25}.  All values in columns 2, 3 and 4 are in units of $10^{-5}$.   
The errors in the second column are the
error without IB corrections, the uncertainty on the IB correction, and the 
total error in square brackets; see the main text.
 The results of Ref.~\cite{Mainz25} have been multiplied by 2 to account for the isospin 
 relation $\r^{I=1}_{e^+e^-}=\half\r_{V}$.  The pull 
is defined as the difference between the lattice and $\t$-based values, normalized
by total error in the 2nd column and the error in the 4th column added in quadrature.}}
\end{table}
%%%%%%%%%

The central values for $\bP(Q^2)$ obtained from the dispersive
representation of Eqs.~(\ref{vacpolQ2}) and~(\ref{vacpolQ2decomp}) are shown in the
second column of the table.   The IB corrections listed in the third column
have already been applied.
The first error on the $\bP(Q^2)$ entries in the second column is that obtained 
by adding in 
quadrature the errors 
on the data, PT and DV parts.
The second error shown is the systematic uncertainty associated with IB 
corrections,
described in Sec.~\ref{isospin} above; this error is also shown in the third column.   
The total error is obtained by summing these
two errors in quadrature, and is shown in square brackets in the second column.   
The fourth column 
shows the corresponding lattice values, obtained by multiplying the results 
of Ref.~\cite{Mainz25} by a factor two to match our
normalization convention.

Finally, the last column of the table shows the tension between the $\t$-based
representation and the results of Ref.~\cite{Mainz25}, quantified as a pull for each 
value of $Q^2$,
obtained by taking the difference and dividing by the quadrature sum of the 
total errors in the 2nd and 4th columns.   We see that there is a $2.1-3.0\s$ tension, 
with the $\t$-based results lying systematically below the lattice values.

The BMW collaboration, in Ref.~\cite{BMW20}, reported a lattice value
\begin{equation}
\label{BMWvalue}
\bP(Q^2=1\ \mbox{GeV}^2)=0.06453(42)\ ,
\end{equation}
where we assigned the finite-volume correction between the reference volume
and
infinite volume entirely to the lqc part, and multiplied by $9/10$ to convert from 
the BMW lqc normalization
to the $I=1$ normalization. The BMW result differs from our 
$1$~GeV$^2$ result by $1.5\s$.   A value for $\bP(Q^2=10\ \mbox{GeV}^2)-
\bP(Q^2=1\ \mbox{GeV}^2)$
was also reported in Ref.~\cite{BMW20}:  
\begin{equation}
\label{BMW10minus1}
\bP(Q^2=10\ \mbox{GeV}^2)-\bP(Q^2=1\ \mbox{GeV}^2)=0.06545(40)\ ,
\end{equation}
where we again adjusted to the $I=1$ normalization used in this paper.  
Our value at $10$~GeV$^2$ is $\bP(10\ \mbox{GeV}^2)=0.12875(68)(46)[82]$,
which includes the IB correction $-0.00057(46)$, in the notation of 
Table~\ref{tab:PiMainz}.
However, without a more detailed understanding of the correlations between the 
systematic uncertainties on our estimates of the IB corrections at different
$s$, the strength of the correlations between the uncertainties on the 
associated IB corrections to $\bP(Q^2)$ at different $Q^2$ is unclear. 
However, one obtains the smallest uncertainty for the IB correction 
to the dispersive version of the difference in Eq.~(\ref{BMW10minus1}) if
one assumes the errors on the IB corrections at 1~GeV$^2$ and 10~GeV$^2$ 
are fully (and positively) correlated. In this case we find a dispersive
result that differs by $0.61\, \sigma$ from the lattice result in 
Eq.~(\ref{BMW10minus1}). If instead we assume the IB errors at
1~GeV$^2$ and 10~GeV$^2$ to be uncorrelated, this difference is 
reduced to $0.44\s$.

From Table~\ref{tab:PiMainz}, it can be seen that the errors obtained for the $\t$-based 
representation are competitive with those obtained on the lattice, especially
for lower $Q^2$ values, where the data part of $\bP(Q^2)$ constitutes a larger
part of the total.   For $Q^2=0.5$~GeV$^2$, the data part is about $91\%$ of the total
and the perturbative part, \ie, the contribution to $\bP(Q^2)$ above $s\approx m_\t^2$
({\it c.f.} Eq.~(\ref{pertpart})) about $9\%$.   For $Q^2=12$~GeV$^2$ these percentages
are about $68\%$ and $32\%$, respectively. For all values of $Q^2$, the DV part is only a fraction of a percent.  In addition, the magnitude of the IB correction, and the 
corresponding systematic error, increase with increasing $Q^2$,
though the size of the IB correction relative to $\bP(Q^2)$ decreases 
with increasing $Q^2$, as one would expect, varying from $0.65\%$ at 
$Q^2=0.5$~GeV$^2$ to $0.43\%$ at $Q^2=12$~GeV$^2$.

\subsection{\label{windows} Light-quark connected window quantities}

In this subsection, we compare the quantities defined in Eqs.~(\ref{amuHVP}) 
and~(\ref{amuW}) obtained from the $\t$-based representation for $C(t)$ of
Eq.~(\ref{Ct}) to the average lattice values provided in Ref.~\cite{WP25}, which is based
on the work of Refs.~\cite{RBC18,BMW20,Mainz24,ExtendedTwistedMass:2022jpw,
RBC:2023pvn,MILC:2024ryz,Giusti:2018mdh,
Shintani:2019wai,Giusti:2019hkz,Lehner:2020crt,
Wang:2022lkq,Aubin:2022hgm,Ce:2022kxy,Boccaletti:2024guq,Spiegel:2024dec,
RBC:2024fic,Djukanovic:2024cmq,ExtendedTwistedMass:2024nyi,
FermilabLatticeHPQCD:2024ppc}.   Figures showing 
the individual results from these references will be shown in Sec.~\ref{exclusive}.

Results are shown in Table~\ref{tab:amu}.   The row ``$\t$-data'' shows
our results, without IB corrections, and with errors obtained analogously 
to those in Table~\ref{tab:PiMainz}.
The row ``WP25'' shows the lattice averages from Ref.~\cite{WP25}, with two 
consistent averages for $a_\m^{\rm HVP,lqc}$ (for explanation of lattice averaging 
procedures, we refer to Ref.~\cite{WP25}).   Our results are quoted using the definition of
isoQCD of Ref.~\cite{FLAG}, which differs slightly from the scheme of Ref.~\cite{WP25}. However, within current errors
the two definitions are sufficiently close for this comparison.

Clearly, the $\t$-based and lattice results
are consistent, with the largest relative difference, which occurs for the SD window,
equal to $1.0\s$.   We note that for the LD window and the full HVP, the $\t$-based
results have somewhat smaller errors, while the $\t$-based SD and W1 values are not 
as precise as the corresponding lattice results.

%%%%%%%%%
\begin{table}[t]
\begin{center}
\begin{tabular}{ccccc}
\hline
Mode & SD & W1 & LD & HVP \\
\hline
$\t$-data & 25.61(28) & 192.36(88) & 406.2(2.5) & 624.2(3.2)\\
PT+DV & 22.26(21) & 14.58(39) & 0.669(38) & 37.52(63) \\
EM IB & $-$0.13(11) & $-$1.40(1.1) & $-$2.2(1.9) & $-3.7(3.1)$ \\
\hline
lqc total & $47.74(37)$ & $205.55(1.5)$ & $404.7(3.1)$ & $657.9(4.5)$ \\
\hline
WP25 & 48.123(83) & 206.97(41) & 406.0(4.9) & 659.5(4.7) (Avg. A) \\
& & & & 661.1(5.0) (Avg. B) \\
\hline
\end{tabular}
\end{center}
\floatcaption{tab:amu}%
{{\it Comparison of $\t$-based quantities $a_\m^{\rm X,lqc}$ for X=SD, W1, LD and HVP 
with the WP25 average values \cite{WP25}.   For the two different lattice HVP averages,
see Ref.~\cite{WP25}.
All entries in units of $10^{-10}$. See the main text.}}
\end{table}
%%%%%%%%%

Since large-$t$ contributions to the SD window is strongly 
suppressed by the corresponding window functions, 
it is interesting to compare the $\t$-based values for the SD 
window quantity with that obtained in PT only, in pure QCD with massless
up and down quarks \cite{CM}.   For the SD window, we find
\begin{equation}
\label{PTonly}
a_\m^{\rm SD,lqc,PT}=47.0(1.2)\times 10^{-10}\ ,
\end{equation}
This value is about $0.5\s$ lower than the 
non-perturbative, $\t$-based value shown in Table~\ref{tab:amu}, with the
perturbative error in Eq.~(\ref{PTonly}) being much larger than that in the 
table.  It is somewhat
surprising that PT gives a good value for the SD window, as that 
window has a significant contribution, about 36\% ({\it c.f.} Table~\ref{tab:windows-tau2pi}), from the $2\p$ mode, which
includes the $\r$ peak region.

\begin{figure}[t!]
\vspace*{4ex}
\begin{center}
\includegraphics*[width=10.25cm]{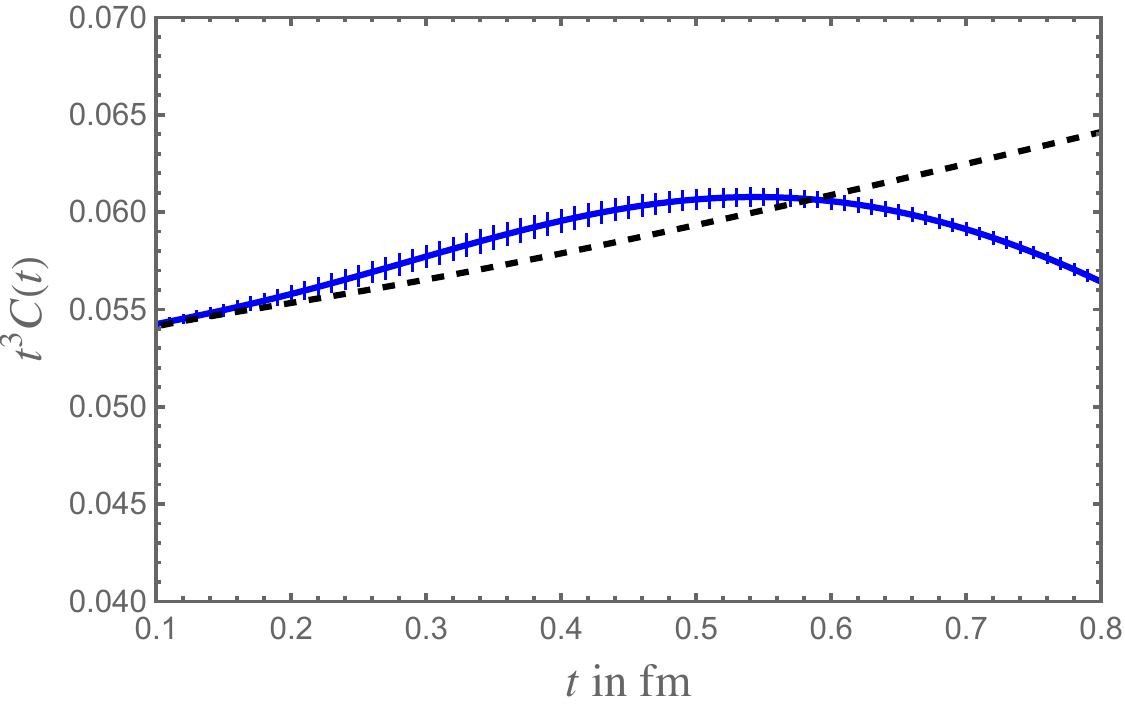}
\end{center}
\begin{quotation}
\floatcaption{fig:comparePT}%
{{\it Comparison of $t^3 C(t)$ between the $\t$-based representation (blue)
and QCD perturbation theory (black,dashed) with $\a_s(m_\t^2)=0.317$.}}
\end{quotation}
\vspace*{-4ex}
\end{figure}

Finally, we compare $C(t)$ from our $\t$-based representation with pure
PT.   As $C(t)\sim 1/t^3$ for $t\to 0$, we show the combination $t^3 C(t)$
in Fig.~\ref{fig:comparePT}, with the $\t$-based representation in blue
and the purely perturbative version in black, in the range between $0.1$
and $0.8$~fm.\footnote{This figure is inspired by Fig.~7 of Ref.~\cite{RBC:2023pvn}.}
Small non-perturbative effects are already visible in the region between 0.2 and
0.6~fm.

\subsection{\label{exclusive} Comparisons based on exclusive \begin{boldmath}$\t$-decay\end{boldmath} mode contributions}

In the previous subsection, the isospin-limit lqc contributions to the RBC/UK\-QCD 
windows and to the total HVP were obtained from the inclusive $I=1$ 
$\tau\to{\rm hadrons}+\nu_\tau$ spectral function, supplemented 
with PT and DV contributions for $s>3.087$~GeV$^2$. In a series of previous papers, 
using the strategy of Refs.~\cite{Boito:2022dry,Boito:2022rkw}, we obtained  data-driven, results for the lqc and the strange-plus-light-quark-disconnected (s+lqd) contributions to the same window quantities and to the total HVP from $e^+e^-$ data~\cite{Benton:2023dci, Benton:2023fcv, Benton:2024kwp}, using pre-CMD-3 combined exclusive-mode KNT19 distributions provided by the authors of~Ref.~\cite{KNT19}.  Results for the lqc and s+lqd contribution to the total HVP~\cite{Boito:2022dry,Boito:2022rkw,Benton:2024kwp} were also  obtained using publicly available electroproduction results from Davier {\it et al}.~\cite{Davier:2019can}. Given the present discrepancies in the $e^+e^-\to \pi^+\pi^-$ cross-section results from different
experiments~\cite{WP25}, an interesting exercise carried out in 
Ref.~\cite{Benton:2024kwp} was to replace the  pre-CMD-3 KNT19 combination 
of $e^+e^-\to \pi^+\pi^-$ cross sections with the recent 
CMD-3 results \cite{CMD3}, where available. We showed that, using the 
new CMD-3 results, the data driven lqc and s+lqd contributions in the isospin limit 
agree well with the lattice results.

It is therefore interesting to re-obtain the lqc contribution to the RBC/UKQCD windows 
and the total HVP by replacing the pre-CMD-3 KNT19  $e^+e^-\to \pi^+\pi^-$ 
combination  with the $\tau$-based $2\pi$ spectral distribution where available 
(\ie, up to $s=3.087$~GeV$^2$), keeping all other contributions from the KNT19 
compilation used in our previous work~\cite{Benton:2024kwp} unchanged. This
allows us to investigate whether using the $\tau$-based $2\pi$ spectral 
distribution leads to agreement between isospin-symmetric data-driven and 
lattice results, as already observed to happen when the new CMD-3 two-pion 
cross-section results were employed. In addition, given the tensions observed in 
Sec.~\ref{fourpi} between the $\t$-based and $e^+e^-$-based four-pion 
mode distributions, 
it is of interest to obtain $\tau$-based results for the lqc windows where both the 
$2\pi$ and the $4\pi$ mode distributions from the KNT19 compilation are 
replaced by their $\tau$-based counterparts in the energy region up to the 
$\t$ mass.

Compared to our previous determinations of the isospin-symmetric lqc windows, 
the $\tau$-based results have the advantage of being free from leading-order strong IB, as already discussed in Sec.~\ref{isospin}.  
In comparing with the inclusive $\tau$-based results of the previous section, 
one difference is in the switch point at which one transitions from 
using data to using PT, supplemented by DV contributions. When 
using  KNT19 data, this switch point occurs at a higher energy, 
$s=s_{\rm KNT}=3.754$~GeV$^2$. Other  small differences  appear in the 
treatment of 
contributions from sub-dominant modes. In the case of the inclusive $\tau$-based 
$I=1$ spectral function, BaBar results for $\tau\to K^-\bar K^0\n_\t$
\cite{babarkkbartau18} 
were used to obtain the $K\bar K$ contribution. For the $K\bar{K}\pi$
mode, where both vector and axial vector contributions exist, the vector
component was obtained using CVC and the 
$I=1$ vector part of the $e^+e^-\rightarrow K\bar{K}\pi$ distribution
produced by BaBar's Dalitz-plot-based separation of the $I=0$ and $I=1$
components~\cite{babarepemkkbarpi}.
 Contributions from all other modes are obtained from CVC-related $e^+e^-$ 
 cross-sections, resorting to a maximally conservative separation, assigning 
 $50\%\pm 50\%$ of the total to each of $I=1$ and $I=0$ for $G$-parity ambiguous 
 modes,  such as $K\bar K2\pi$. In the evaluation based on $e^+e^-$  cross-sections, 
 apart from the modes already considered in the construction of the inclusive 
 $I=1$ $\tau$-based spectral function, the higher switch point means that 
small contributions from additional modes such as $e^+e^-\to p \bar p$ 
and  $e^+e^-\to n \bar n$, having ambiguous isospin and present only at 
higher $s$, have to be included as well. For these ambiguous modes the 
maximally conservative isospin separation is again used~\cite{Boito:2022dry, 
Boito:2022rkw}.

In Table~\ref{tab:windows-tau2pi}, we show the results where only the $2\pi$ 
KNT19 contribution is replaced by the contribution from the combined 
$\tau\to \pi^-\pi^0\nu_\tau$ spectral distribution. Since the input $\tau$ spectral 
function ends at $s=3.087$~GeV$^2$ a small contribution from the region between 
$3.087$~GeV$^2$  and the KNT19 exclusive-mode
endpoint, $s_{\rm KNT}=3.754$~GeV$^2$, obtained using KNT19 results, has to be added. 
The contributions from all other modes are obtained exactly as in 
Ref.~\cite{Benton:2024kwp} and include the contributions from all
$G=+1$, $G$-parity unambiguous, non-2$\pi$ modes, and our assessments of the 
vector channel parts of all $G$-parity-ambiguous-mode contributions, 
supplemented by PT and DV contributions above the KNT19 endpoint. The EM IB 
correction is the same as the one shown in Table~\ref{tab:amu}. Comparing these 
results with the WP25 averages shown in Table~\ref{tab:amu}, we see that the 
replacement of (pre-CMD-3) KNT19 $2\pi$ input with 
$\tau$-based $2\pi$ input leads to a good agreement with the lattice results
for all four quantities considered.

Next, we replace the KNT19 input for the $2\pi$ and two $4\pi$ modes 
with that obtained using our combined $\tau\to\pi^-\pi^0\nu_\tau$, 
$\tau\to \pi^-3\pi^0\nu_\tau$, and $\tau\to 2\pi^-\pi^+\pi^0\nu_\tau$ spectral 
distributions \cite{us25}. The results of this exercise are shown in 
Table~\ref{tab:windows-tau2pi4pi} and are again compatible, within errors, with the 
WP25 lattice averages.
The values shown in Tables~\ref{tab:windows-tau2pi} and~\ref{tab:windows-tau2pi4pi} 
share several common ingredients and are, therefore, positively correlated.  
This is especially true for the LD window, for which
more than $98\%$ of the total comes from the $2\pi$ mode. Small 
differences between the two strategies are thus potentially significant.
This includes the discrepancies found in Sec.~\ref{fourpi} between the 
$\tau$ and $e^+e^-$ $4\pi$-mode contributions. We estimate the correlation 
between the results for the LD window in 
Table~\ref{tab:windows-tau2pi} and that in Table~\ref{tab:windows-tau2pi4pi} to 
be $99.66\%$, which makes the two results incompatible at the $3.2\sigma$ level. 
The other windows are compatible within $2.3\sigma$ or better.  Results obtained 
with the inclusive $I=1$ spectral function, Table~\ref{tab:amu}, lie between those 
of Tables~\ref{tab:windows-tau2pi} and~\ref{tab:windows-tau2pi4pi} and are 
compatible with them, even in the presence of strong correlations (especially for 
the LD window).

\begin{table}[t]
\begin{center}
\begin{tabular}{lcccc}
\hline
Mode                             & SD          & W1         & LD           & HVP \\
\hline
$\tau$-based $2\pi$              & 17.152(66)  & 165.93(65) & 398.5(2.5)   & 581.6(3.0)  \\
KNT19 $2\pi$ tail                & 0.0535(40)  & 0.0964(72) & 0.00897(67)  & 0.159(12)   \\
Unamb. non 2$\pi$                & 9.41(21)    & 26.77(58)  & 7.02(17)     & 43.19(95)   \\
Amb. modes                       & 0.75(32)    & 1.88(61)   & 0.613(91)    & 3.2(1.0)    \\
PT+DV                       & 20.28(10)   & 11.06(16)  & 0.346(11)    & 31.68(28)   \\
EM IB                            & $-0.13(11)$ & $-1.4(1.1)$ & $-2.2(1.9)$ & $-3.7(3.1)$\\
\hline
lqc total                        & 47.51(42)   & 204.3(1.5) & 404.3(3.1)   & 656.2(4.5)  \\
\hline
\end{tabular}
\end{center}
\floatcaption{tab:windows-tau2pi}%
{{\it Isospin symmetric lqc contribution to the  RBC/UKQCD windows and to the total 
HVP. Here, the $2\pi$ mode contributions are obtained from our combined 
$\tau^-\to\pi^-\pi^0\nu_\tau$ spectral distribution~\cite{us25} (``$\tau$-based $2\pi$''), 
while KNT19 input~\cite{KNT19}, as in 
Ref.~\cite{Benton:2024kwp}, is used for the contributions from all other modes. ``KNT19 
$2\pi$ tail'' refers to the $2\pi$ contribution from the region of $s$ between 
$3.087$~{\rm GeV}$^2$  and $s_{\rm KNT}$, while ``Unamb.'' and ``Amb.'' 
refer to the sums of $G$-parity unambiguous- and ambiguous-mode 
distributions, respectively. Results in units of $10^{-10}$.}}
\end{table}

\begin{table}[t]
    \begin{center}
    \begin{tabular}{lllll}
        \hline
        Mode  & SD & W1 & LD  & HVP \\
        \hline
        $\tau$-based 2$\pi$+4$\pi$ & 24.49(28) & 189.17(87) & 405.4(2.5) & 619.1(3.2) \\
        KNT19 $2\pi$ tail & 0.0535(40) & 0.0964(72) & 0.0090(67) & 0.159(12) \\
        KNT19 $4\pi$ tail & 1.177(24) & 2.122(44) & 0.1981(42) & 3.498(73) \\
        Unamb. non 2$\pi$/4$\pi$ & 1.383(98) & 3.17(19) & 0.578(23) & 5.13(32) \\
        Amb. modes               & 0.75(32)    & 1.88(61)   & 0.613(91)    & 3.2(1.0)    \\
        PT+DV & 20.28(10) & 11.06(16) & 0.346(11) & 31.68(28) \\
        EM IB    & $-0.13(11)$ & $-1.4(1.1)$ & $-2.2(1.9)$ & $-3.7(3.1)$\\
        \hline
        lqc total  & 48.00(46)  & 206.1(1.5) & 405.0(3.1) & 659.1(4.5)  \\
        \hline
    \end{tabular}
    \end{center}
        \floatcaption{tab:windows-tau2pi4pi}%
{{\it Isospin symmetric lqc contribution to the  RBC/UKQCD windows and to the total 
HVP. Here, the $2\pi$ and the $4\pi$ mode contributions are obtained from our 
combined $\tau\to\pi^-\pi^0\nu_\tau$, $\tau\to \pi^-3\pi^0\nu_\tau$, and 
$\tau\to 2\pi^-\pi^+\pi^0\nu_\tau$ spectral distributions~\cite{us25} 
(``$\tau$-based $2\pi+4\pi$''), while contributions from all other modes are 
obtained using 
combined $e^+e^-$ KNT19 input~\cite{KNT19}, as in Ref.~\cite{Benton:2024kwp}.
``KNT19 $2\pi$ tail'' and ``KNT19 $4\pi$ tail''  refer to the $2\pi$-  and $4\pi$-mode 
contributions from the region of $s$ between $3.087$~{\rm GeV}$^2$  and 
$s_{\rm KNT}$.
     ``Unamb.'' and ``Amb.'' refer to the sums of $G$-parity unambiguous- 
     and ambiguous-mode contributions, respectively. Results in units of $10^{-10}$.}}
\end{table}

In Fig.~\ref{fig:lqc-windows}, we show the lqc results from 
Table~\ref{tab:windows-tau2pi4pi} compared with several recent lattice determinations, 
the WP25 averages (grey bands), as well as the previously obtained $e^+e^-$-based 
determinations of the lqc windows~\cite{Benton:2024kwp}. It is clear that the 
pre-CMD-3 $e^+e^-$-based results (in red) are in poor agreement with lattice 
results, 
especially for the W1 window, while those produced using either CMD-3 
$e^+e^-$ $2\pi$ input (in blue) 
or our new $\tau$-based input (in purple) are all in good agreement with the lattice 
determinations of the same quantities.

\begin{figure}[!ht]
\begin{center}
\includegraphics[width=0.49\textwidth]{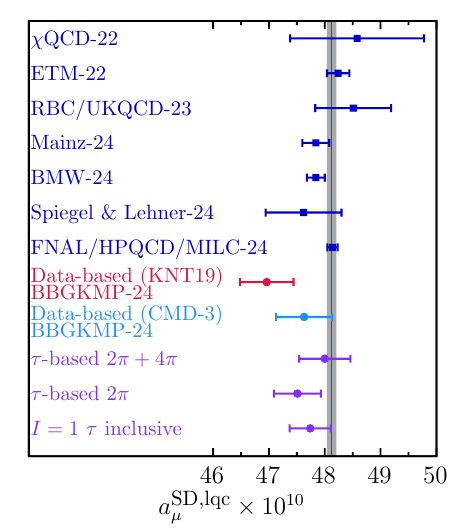}
\includegraphics[width=0.49\textwidth]{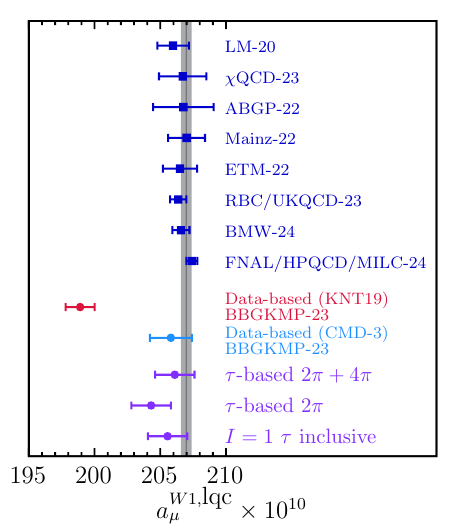}
\includegraphics[width=0.49\textwidth]{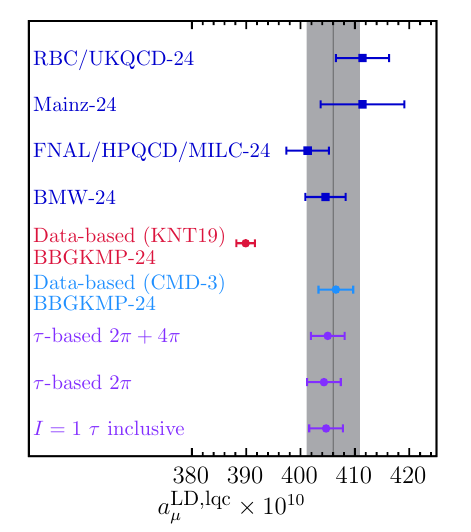}
\includegraphics[width=0.49\textwidth]{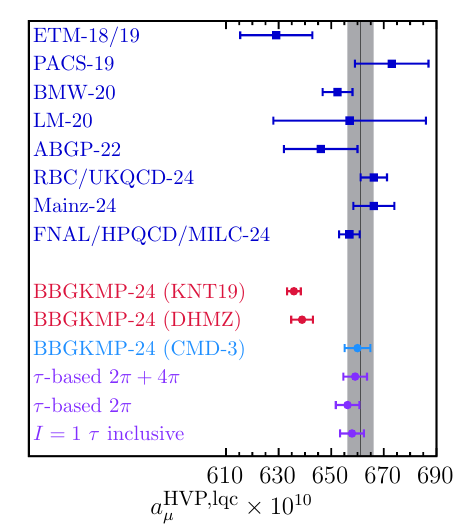}
\end{center}
\floatcaption{fig:lqc-windows}%
{{\it Data-driven results for isospin-symmetric lqc contribution to the RBC/UKQCD 
windows and to the total HVP compared with lattice determinations from 
Refs.~\cite{BMW20,Mainz24,ExtendedTwistedMass:2022jpw,RBC:2023pvn,
MILC:2024ryz,Giusti:2018mdh,Shintani:2019wai,Giusti:2019hkz,Lehner:2020crt,
Wang:2022lkq,Aubin:2022hgm,Ce:2022kxy,Boccaletti:2024guq,Spiegel:2024dec,
RBC:2024fic,Djukanovic:2024cmq,FermilabLatticeHPQCD:2024ppc} and the WP25 
lattice averages (grey bands).
Results based exclusively on $e^+e^-$ data are 
labeled ``BBGKMP-24''~\cite{Benton:2024kwp}. The $\tau$-based results
correspond to those of Tables \ref{tab:amu},~\ref{tab:windows-tau2pi} and~\ref{tab:windows-tau2pi4pi}.}}
\end{figure}

\section{\label{conclusion} Conclusion}
In this paper, we considered the isospin-1 component of the HVP,
which is relevant for high-precision computations of quantities such as the muon 
anomalous magnetic moment and the hadronic contribution
$\D\a_{\rm had}$ to the running of the EM coupling $\a$.    We compared
values obtained dispersively using $I=1$ hadronic $\t$-decay data with those 
obtained in lattice QCD.   Specifically, we considered the subtracted HVP $\bP(Q^2)$, 
as well as the lqc contributions to the SD, W1 and LD window quantities and 
$a_\m^{\rm HVP}$ itself.    As shown in 
Fig.~\ref{fig:lqc-windows}, generally good agreement was found between 
the $a_\mu^{\rm HVP,lqc}$ and lqc window quantities produced using the 
$\t$-based representation and lattice results for these same quantitities 
reported and averaged in Ref.~\cite{WP25}. We note that lattice 
errors are smaller for the SD and W1 window quantities, while
the situation is reversed for the LD window and $a_\m^{\rm HVP,lqc}$.
For $\bP(Q^2)$, with $0.5$~GeV$^2\le Q^2\le 12$~GeV$^2$, we found 
$\t$-based values which lie $2-3\s$ below those of Ref.~\cite{Mainz25}, and 
are in better agreement with the smaller number of results reported 
in Ref.~\cite{BMW20}.

In order to make the comparisons, the $\t$-data-based quantities had to be corrected
for IB, which we assumed to be dominated by those of the two-pion mode.   While 
radiative corrections appear to be under control \cite{CCHH}, IB also affects the
form factor $f_+(s)$, for which no model-independent analysis of IB effects exists
at present.   For this form factor, we relied on the ChPT-inspired model of Ref.~\cite{CEN} 
with what we believe to be conservative errors,
following recommendations from Ref.~\cite{WP25}.

Finally, we compared the four-pion spectral distributions obtained
using $\t$ data with expectations for those distributions produced
using electroproduction data and the isospin-based Pais relations.  
Particularly interesting comparisons based on these results are
between the four-pion contributions to the
SD, W1 and LD window quantities, where we find significant differences between the
$\t$-based $2\p^-\p^+\p^0$ exclusive-mode contributions 
and their $e^+e^-$-based counterparts.
These differences do not readily show up in our all-contribution totals for these 
lqc quantities, as the
$2\p^-\p^+\p^0$ differences, though themselves statistically 
significant when all correlations are taken into account, are smaller than the errors 
on these totals.

\vspace{3ex}
\noindent {\bf Acknowledgments}
\vspace{3ex}

DB and MG would like to thank Andreas Maier, MG and SP Pere Masjuan and MG   
Vincenzo Cirigliano and Alessandro Conigli for discussions. DB would like to
thank the IFAE and UAB for hospitality.  NA and MG are supported by the 
U.S.\ Department of Energy, Office of Science, Office of High Energy 
Physics, under Award No.~DE-SC0013682.  MG is also supported as a
Severa Ochoa visitor to the IFAE.  DB's work was supported by the S\~ao Paulo 
Research Foundation (FAPESP) grant No. 2021/06756-6 and by CNPq grant 
No. 303553/2025-1.  KM is supported by a grant from the 
Natural Sciences and Engineering Research Council of Canada.
LMM is supported by the Deutsche Forschungsgemeinschaft (DFG, German 
Research Foundation)--project 
number: 514321794 (CRC1660: Hadron and Nuclei as discovery tools).
SP is supported by the Spanish Ministerio de Ciencia e Innovacion,
grants PID2020-112965GB-I00, PID2023-146142NB-I00, under
the Grant CEX2024-001441-S funded by MICIU/AEI/10.13039/501100011033,
 and by the Departament de Recerca i Universitats from Generalitat de Catalunya
 to the Grup de Recerca 00649 (Codi: 2021 SGR 00649).
IFAE is partially funded by the CERCA program of the Generalitat de Catalunya. 

\appendix
\section{\label{rhoder} Derivation of Eq.~(\ref{rhoPT})}

We start from the (unsubtracted) vacuum polarization as a function of $s$,
\begin{equation}
\label{spec1}
\P(s)=-\frac{1}{4\p^2}\left(c_{00}+\log\frac{-s}{\m^2}+
\sum_{n=1}^\infty a^n(\m^2)\sum_{m=0}^n c_{nm}\log^m\frac{-s}{\m^2}\right)\ ,
\end{equation}
where $a(\m^2)=\a_s(\m^2)/\p$ and the independent coefficients,
$c_{n1}$, are given in Eq.~(\ref{cs}). The values of the remaining $c_{nm}$, 
$m>1$, which follow from RG considerations, are given in Eq.~(\ref{othercs}) below.
The spectral function is defined as ($\e>0$)
\begin{equation}
\label{specdef}
\r(s)=\lim_{\e\to 0}\frac{1}{2\p i}\left(\P(s+i\e)-\P(s-i\e)\right)\ .
\end{equation}
Using the fact that $\P(s)$ is analytic in the complex plane except on the positive
real axis, and setting $\m^2=s$, this can be written as
\begin{eqnarray}
\label{spec2}
\r(s)&=&-\frac{1}{2\p i}\oint_{|z|=s}dz\,\frac{d\P(z)}{dz}\\
&=&\frac{1}{8\p^3 i}\oint_{|x|=1}\frac{dx}{x}\left(1+\sum_{n=1}^\infty 
a^n(s)\sum_{m=1}^n mc_{nm}\log^{m-1}(-x)\right)\nonumber\\
&=&\frac{1}{4\p^2}\left(1+\sum_{n=1}^\infty a^n(s)\sum_{m=1}^n c_{nm}J_m\right)
\nonumber
\nonumber
\end{eqnarray}
using
\begin{equation}
\label{Jint}
\frac{m}{2\p i}\oint_{|x|=1}\frac{dx}{x}\log^{m-1}(-x)=J_m\ ,\quad m>0\ ,\ \mbox{integer}\ ,
\end{equation}
with $J_m$ given in Eq.~(\ref{Js}).

For completeness,
we list the values of $c_{nm}$ for $n\ge m>1$:
\begin{eqnarray}
\label{othercs}
c_{22}&=&-1.125\ ,\\
c_{32}&=&-5.6896\ ,\quad\ \,c_{33}=1.6875\ ,\nonumber\\
c_{42}&=&-33.0914\ ,\quad c_{43}=15.8016\ ,\quad c_{44}=-2.84766\ ,\nonumber\\
c_{52}&=&-299.177\ ,\quad c_{53}=129.578\ ,\quad c_{54}=-40.6161\ ,\quad
c_{55}=5.12578\ .
\nonumber
\end{eqnarray}

\section{\label{formulas} Formulas}

The one-loop function $h_{ij}(s,\m)$ is given by
\begin{eqnarray}
\label{h}
h_{ij}(s,\m)&=&\frac{1}{12s}\,\l(s,m_i^2,m_j^2)\overline{J}(s,m_1,m_2)
+\frac{1}{18(4\p)^2}\,(s-3(m_i^2+m_j^2))\\
&&-\frac{1}{12}\left(\frac{2(m_i^2+m_j^2)-s}{m_i^2-m_j^2}(L_i(\m)-L_j(\m))-2(L_i(\m)
+L_j(\m))\right)\ ,
\nonumber
\end{eqnarray}
where
\begin{equation}
\label{L}
L_i(\m)=-\frac{m_i^2}{(4\p)^2}\,\log\frac{m_i^2}{\m^2}\ ,
\end{equation}
and
\begin{eqnarray}
\label{Jbar}
\overline{J}(s,m_1,m_2)&=&
-\frac{1}{16\p^2}\int_0^1 dx\log\frac{(1-x)m_1^2+xm_2^2-x(1-x)s}{(1-x)m_1^2+
xm_2^2}\\
&=&\frac{1}{32\p^2}\left(2+\left(\frac{m_1^2-m_2^2}{s}-\frac{m_1^2+m_2^2}{m_1^2-
m_2^2}\right)\log\frac{m_2^2}{m_1^2}+\frac{\l^{1/2}(s,m_1^2,m_2^2)}{s}
\times\right.\nonumber\\
&&\hspace{-2.5cm}\left.\left(\log\frac{m_1^2-m_2^2-s+
\l^{1/2}(s,m_1^2,m_2^2)}{\l^{1/2}(s,m_1^2,m_2^2)}+\log\frac{-m_1^2+m_2^2-s+
\l^{1/2}(s,m_1^2,m_2^2)}{\l^{1/2}(s,m_1^2,m_2^2)}\right.\right.\nonumber\\
&&\hspace{-2.5cm}\left.\left.-\log\frac{m_1^2-m_2^2+s+
\l^{1/2}(s,m_1^2,m_2^2)}{\l^{1/2}(s,m_1^2,m_2^2)}-\log\frac{-m_1^2+m_2^2+s
+\l^{1/2}(s,m_1^2,m_2^2)}{\l^{1/2}(s,m_1^2,m_2^2)}\right)\right)
\ .\nonumber
\end{eqnarray}
One has to be careful with combining the logarithms in this expression, as both the
numerators and denominators can have complex phases.
Here the logarithm is defined to have a branch cut discontinuity from $-\infty$ to 0, and
the expressions above need to be evaluated at $s+i\e$ with $\e>0$ in the limit $\e\to 0$.

%%%%%%%%%%%%%%%%%%%%%%%%%%%

\end{document}